%% file: ms.tex
\documentclass[preprint]{aastex}

\shorttitle{M, L, and T Sequence}
\shortauthors{Cushing et al.}

\received{2004 October 4}
\begin{document}


\title{An Infrared Spectroscopic Sequence of M, L and T Dwarfs\altaffilmark{1}}


\author{Michael C. Cushing\altaffilmark{2,3}}
\affil{SETI Institute, MS 245-3 NASA Ames Research Center, Moffett Field, CA 94035}
\email{mcushing@mail.arc.nasa.gov}

\author{John T. Rayner}
\affil{Institute for Astronomy, University of Hawai`i, 2680 Woodlawn Drive, Honolulu, HI 96822}
\email{rayner@ifa.hawaii.edu}

\and

\author{William D. Vacca\altaffilmark{2}}
\affil{SOFIA-USRA, NASA Ames Research Center MS 144-2, Moffett Field, CA 94035} 
\email{wvacca@mail.arc.nasa.gov}

\altaffiltext{1}{Based in part on data collected at Subaru Telescope, which 
  is operated by the National Astronomical Observatory of Japan.}

\altaffiltext{2}{Visiting Astronomer at the Infrared Telescope Facility, which is operated by the University of Hawai`i under cooperative Agreement no. NCC 5-538 with the National Aeronautics and Space Administration, Office of Space Science, Planetary Astronomy Program.}
\altaffiltext{3}{Spitzer Fellow}

\begin{abstract}

We present a 0.6$-$4.1 $\mu$m spectroscopic sequence of M, L, and T dwarfs.  The spectra have  $R$ $\equiv$ $\lambda/\Delta\lambda$ $\approx$ 2000 from 0.9 to 2.4 $\mu$m and $R$=2500$-$200 from 2.9 to 4.1 $\mu$m.  These new data nearly double the number of L and T dwarfs that have reported $L$-band spectra.  The near-infrared spectra are combined with previously published red-optical spectra to extend the wavelength coverage to $\sim$0.6 $\mu$m.  Prominent atomic and molecular absorption features are identified including neutral lines of Al, Fe, Mg, Ca, Ti, Na, and K and 19 new weak CH$_4$ absorption features in the $H$-band spectra of mid- to late-type T dwarfs.  In addition, we detect for the first time the 0$-$0 band of the $A$ $^4\Pi- X$ $^4\Sigma^-$ transition of VO at $\sim$1.06 $\mu$m in the spectra of L dwarfs and the P and R branches of the $\nu_3$ band of CH$_4$ in the spectrum of a T dwarf.  The equivalent widths of the refractory atomic features all decrease with increasing spectral type and are absent by a spectral type of $\sim$L0, except for the 1.189 $\mu$m Fe I line which persists to at least $\sim$L3.  We compute the bolometric luminosities of the dwarfs in our sample with measured parallaxes and find good agreement with previously published results that use $L'$-band photometry to account for the flux emitted from 2.5 to 3.6 $\mu$m.  Finally, 2MASS J2224381$-$0158521 (L4.5) has an anomalously red spectrum and the strongest $\Delta \nu$=$+$2 CO bands in our sample.  This may be indicative of unusually thick condensate clouds and/or low surface gravity.

\end{abstract}


\section{Introduction}\label{Introduction}

Since the discovery of the first bona fide brown dwarfs (BDs) \citep{1996ApJ...458..600B,1996ApJ...469L..53R,1995Natur.378..463N}, hundreds of very low-mass stars and BDs, collectively known as ultracool dwarfs, have been discovered in wide-field surveys such as the 2 Micron All Sky Survey \citep[2MASS;][]{1997ilsn.proc...25S}, the Deep Near Infrared Southern Sky Survey \citep[DENIS;][]{1997Msngr..87...27E}, and the Sloan Digital Sky Survey \citep[SDSS;][]{2000AJ....120.1579Y}.  The spectra of many of these ultracool dwarfs were sufficiently distinct from those of the latest M dwarfs that the creation of two new spectral types, `L' \citep{1999ApJ...519..802K,1999AJ....118.2466M} and `T' \citep{2001PhDT.......116B,2002ApJ...564..421B}, was warranted.  The spectra of L dwarfs exhibit weak oxide bands (TiO and VO), strong hydride bands (FeH and CrH), and alkali lines (Na, K, Rb, and Cs) in the red optical and H$_2$O, FeH, and CO absorption bands in the near-infrared.  The spectra of T dwarfs also show absorption due to alkali lines in the optical (although the Na and K lines are highly pressure-broadened) but, in contrast to the L dwarfs, exhibit strong CH$_4$ bands along with H$_2$O bands and collision-induced absorption (CIA) due to H$_2$ in the near-infrared.

Since the spectral energy distributions of ultracool dwarfs peak in the near-infrared, it is advantageous to study them at these wavelengths.  There have been numerous near-infrared spectroscopic studies of M, L, and T dwarfs \citep[e.g.,][]{1994MNRAS.267..413J,1995AJ....110.2415A,1996ApJS..104..117L,2001AJ....121.1710R,2001ApJ...548..908L,2000AJ....120.1100B,2002ApJ...564..421B} but the majority of the published spectra have low spectral resolving powers ($R$ $\equiv$ $\lambda/\Delta\lambda$ $\lesssim$ 500) and a range of signal-to-noise ratios (S/N).  Although such spectra are suitable for studying variations in gross spectral morphology \citep[e.g. defining spectral classification schemes, ][]{2001AJ....121.1710R,2002ApJ...564..466G,2002ApJ...564..421B}, higher resolution spectra are required for more detailed analyses.  For example, \citet{2000ApJ...541..374S} analyzed an $R$=3000 spectrum of Gl 229B (T6.5) to constrain its metallicity while \citet{2003ApJ...582.1066C} identified nearly 100 new weak FeH absorption features in $R$$\approx$2000 spectra of late-type M dwarfs and L dwarfs.  \citet{2003ApJ...596..561M} have recently published a large sample of high quality, moderate resolution ($R$=2000), near-infrared ($\sim$1$-$2.3 $\mu$m) spectra of late-type M, L, and T dwarfs, which greatly enhances the quality and number of ultracool dwarf spectra available for such analyses.

In contrast to the large number of published spectra covering from 1 to 2.5 $\mu$m, relatively little spectroscopic work on M, L, and T dwarfs has been done at $\lambda > 2.5 \mu$m, primarily due to the difficulty of observing from the ground at these wavelengths.  \citet{1987MNRAS.227..315B} were the first to detect H$_2$O absorption at $\sim$3 $\mu$m in very low resolution ($R$ $\approx$ 70) spectra of early- to mid-type M dwarfs.  \citet{1995MNRAS.277..767J} and \citet{2002MNRAS.330..675J} used moderate resolution spectra obtained from the ground and space to study this H$_2$O band in a small sample of M dwarf spectra.  Even fewer spectra of L and T dwarfs at $\lambda >$ 2.5 $\mu$m have been published.  The $\nu_3$ fundamental band of CH$_4$ at $\sim$3.3 $\mu$m has been detected in the low resolution spectra of two L dwarfs \citep{2000ApJ...541L..75N} and two T dwarfs \citep{1998ApJ...502..932O,2001PhDT.......116B} while the fundamental CO band at 4.7 $\mu$m has been detected in the spectrum of Gl 229B \citep{1997ApJ...489L..87N}.  Recent observations from the \textit{Spitzer Space Telescope} \citep{2004ApJS..154..418R} have also revealed H$_2$O, CH$_4$, and NH$_3$ absorption in the mid-infrared spectra of an L dwarf and a T dwarf.  The paucity of M, L, and T dwarf spectra at these wavelengths is unfortunate since the H$_2$O, CH$_4$, CO, and NH$_3$ bands are useful probes of the atmospheric physics and chemistry of ultracool dwarfs \citep{saumon03,2003IAUS..211..345S}.

We have therefore undertaken a spectroscopic survey of M, L, and T dwarfs with the goal of creating a sample of high S/N ($>$50), moderate resolution ($R$ $\approx$ 2000) spectra covering a broad wavelength range ($\sim$0.9 to 4.1 $\mu$m).  We present the first results of this survey in this paper.  In \S\ref{Observations and Data Reduction} we discuss the observations and data reduction.  We describe the identification of prominent atomic and molecular absorption features as well as quantify the changes in the strengths of these features with spectral type in \S\ref{Analysis}.  In \S\ref{Individual Objects of Interest}, we discuss two objects of interest;  2MASS 2224$-$0158 (L4.5), which exhibits an anomalously red near-infrared spectrum with deep CO bands and DENIS 0255$-$4700, a late-type L dwarf (L8) that exhibits CH$_4$ absorption in both the $H$ and $K$ bands.  In \S\ref{Bolometric Fluxes and Luminosities} we compute the bolometric luminosities of the ultracool dwarfs in our sample and finally in \S\ref{Summary} we present a summary of our results.

\section{Observations and Data Reduction}\label{Observations and Data Reduction}

Our sample consists of 11 M dwarfs, 13 L dwarfs, and 2 T dwarfs.  Table \ref{sample} lists the dwarfs' designations, spectral types, $J$-, $H$-, $K$-, and $L'$-band magnitudes, and parallaxes.  Eight of the M dwarfs were selected from the list of primary spectral standards of \citet{1991ApJS...77..417K}; we have augmented this list with Gl 388 (AD Leo), LP 944$-$20, and BRI 0021$-$0214.  The M dwarf spectral types are based on optical spectroscopy and are from \citet{1991ApJS...77..417K,1995AJ....109..797K,1999ApJ...519..802K}.  The L and T dwarfs were drawn from the discoveries in 2MASS \citep{1999ApJ...519..802K,2000AJ....120..447K,2000AJ....119..369R,2000AJ....120.1085G,2000AJ....120.1100B,2003IAUS..211..197W}, DENIS \citep{1999AJ....118.2466M}, and SDSS \citep{2000ApJ...536L..35L,2000AJ....119..928F}.  We also included Kelu$-$1 which was discovered in a proper-motion survey \citep{1997ApJ...491L.107R}.  The L dwarf spectral types are based on optical spectroscopy and are from the L dwarf database maintained by J.~D. Kirkpatrick\footnote{\url{http://spider.ipac.caltech.edu/staff/davy/ARCHIVE/index\_l\_spec.html}}.  The T dwarf spectral types are based on near-infrared spectroscopy and are from \citet{2002ApJ...564..421B}.

We obtained near-infrared spectra of the 26 dwarfs listed in Table 1 using two instruments: SpeX \citep{2003PASP..115..362R}, mounted on the 3.0 m NASA Infrared Telescope Facility (IRTF), and the Infrared Camera and Spectrograph \citep[IRCS,][]{kobayashi00}, mounted on the 8.2 m Subaru telescope.  SpeX was used for the majority of the observations while the IRCS was used to obtain $L$-band spectra for the faintest targets.  The following three sub-sections describe the acquisition and reduction of the data, the absolute flux calibration of the spectra, and the merging of the SpeX, IRCS, and published red-optical spectra to produce absolutely flux-calibrated 0.6$-$4.1 $\mu$m spectra.

\subsection{SpeX Observations}\label{SpeX Observations}

SpeX is a 0.8$-$5.5 $\mu$m, medium-resolution, cross-dispersed spectrograph equipped with a 1024 $\times$ 1024 Aladdin 3 InSb array \citep{2003PASP..115..362R}.  The entire 0.8 to 5.5 $\mu$m wavelength range can be covered at $R \approx 2000$ with two cross-dispersed modes.  The SXD  mode provides simultaneous coverage of the 0.8$-$2.4 $\mu$m wavelength range, except for a 0.6 $\mu$m gap between the $H$ and $K$ bands, while the LXD1.9, LXD2.1, and LXD2.3 modes cover the 1.9$-$4.2, 2.1$-$5.0 and 2.3$-$5.5 $\mu$m wavelength ranges, respectively.  The length of the slit in all of these modes is 15$\arcsec$.

The observations were conducted over a period of three years beginning with SpeX commissioning in 2000 May and ending in 2003 October.  A log of the observations, including the UT date of observation, spectroscopic mode, resolving power, total on-source integration time, and associated telluric standard star, is given in Table \ref{spexlog}.  The 0$\farcs$3 and 0$\farcs$5 slits were used in the SXD mode and provide a resolving power of 2000 and 1200, respectively.  Three slits with widths of 0$\farcs$3, 0$\farcs$5, and 0$\farcs$8, corresponding to a resolving power of 2500, 1515, and 938, respectively, were used in the LXD modes.

To facilitate subtraction of the additive components of the total signal (electronic bias level, dark current, sky and background emission), the observations were obtained in a series of exposures in which the target was placed at two different positions along the slit separated by 7$\farcs$8.  An A0 V star was observed before or after each target to correct for absorption due to the Earth's atmosphere and to flux calibrate the science object spectra.  The airmass difference between the object and ``telluric standard'' was almost always less than 0.1 and usually less than 0.05.  However, in a few cases where there was a paucity of nearby A0 V stars, the airmass difference was as large as 0.2.  Finally, a set of flat field exposures and argon arc lamp exposures was taken after each object/standard pair for flat fielding and wavelength calibration purposes.

We reduced the data using Spextool \citep{2004PASP..116..362C}, the facility IDL-based data reduction package for SpeX.  The initial image processing consisted of correcting each science frame for non-linearity, subtracting the pairs of images taken at the two different slit positions and dividing the pair-subtracted images by a normalized flat field.  The spectra were then optimally extracted and wavelength calibrated.  All wavelengths are given in vacuum.  The dwarf spectra were corrected for telluric absorption and flux calibrated (to $\pm\sim$ 10\%) using the extracted A0 V spectra and the technique described by \citet{2003PASP..115..389V}.  For each cross-dispersed mode, the spectra from different orders were  merged together to form a continuous spectrum.  In principle, the flux density levels of the spectra in two overlapping orders should match exactly but in practice we find mismatches of up to 3\% but typically less than 1\%.  Any mismatch was removed by scaling one spectrum to the level of the other.  The combined spectra were then smoothed with a Savitzky-Golay kernel \citep{1992nrfa.book.....P} whose width was 1.5 times the slit width measured in pixels.  Each LXD spectrum was scaled to match the flux density level of the corresponding SXD spectrum and then the two spectra were combined.

Although the spectra in adjacent orders are scaled to a common flux density level during the merging process, errors can be introduced in the relative flux density levels of the spectra if the S/N in the overlap regions, which is used to determined the scaling factor, is low.  Such is the case for dwarf stars with spectral types $\ga$M6 because of the strong intrinsic water absorption bands present at the wavelengths of the order overlaps.  Therefore we computed synthetic $J-H$, $H-K$, and $K-L'$ colors of the dwarfs from our sample and compared them with the colors derived from the published photometry presented in Table \ref{sample} to determine if any errors were introduced during the merging process.

The synthetic color for two bandpasses $X$ and $Y$ is given by,

\begin{equation}
X-Y = -2.5 \times \log \left [ \frac{\int \lambda f^{\mathrm{obj}}_{\lambda}(\lambda)T_X(\lambda)d\lambda}{\int \lambda f^{\mathrm{Vega}}_{\lambda}(\lambda)T_X(\lambda)d\lambda} \right ] +2.5 \times \log \left [ \frac{\int \lambda f^{\mathrm{obj}}_{\lambda}(\lambda)T_Y(\lambda)d\lambda}{\int \lambda f^{\mathrm{Vega}}_{\lambda}(\lambda)T_Y(\lambda)d\lambda} \right ] \label{coloreqn}
\end{equation}
\noindent
where $f^{\mathrm{Vega}}_{\lambda}(\lambda)$ is the flux density of Vega, $f^{\mathrm{obj}}_{\lambda}(\lambda)$ is the flux density of the object, and $T(\lambda)$ is the transmission function for each bandpass, which we assume to be given by the product of the filter transmission and the typical atmospheric transmission at an airmass of 1 for the site at which the observation was conducted.  For $f_{\lambda}^{\mathrm{Vega}} (\lambda)$ we used a Kurucz model of Vega ($T_{\mathrm{eff}}=9550$ K, $\log g = 3.950$, $v_{\mathrm{rot}} = 25$ km s$^{-1}$, and $v_{\mathrm{turb}} = 2$ km s$^{-1}$), scaled to the flux density at $\lambda$=5556 \AA \ given by \citet{1995A&A...296..771M}.  Equation \ref{coloreqn} assumes that Vega has a color of zero.  The factors of $\lambda$ inside the integrals convert the energy flux densities $f_{\lambda}$ to photon flux densities, which ensures that the integrated fluxes are proportional to the observed photon count-rate \citep[e.g.,][]{1986HiA.....7..833K,1992A&A...264..557B}.  The CIT $JHK$, and MKO $L'$ transmission functions were kindly provided by S. Leggett (2002 \& 2003,  private communication) while the 2MASS filter transmissions are from \citet{2003AJ....126.1090C}\footnote{Note that the $\lambda$ factors should be removed from Equation \ref{coloreqn} when using the 2MASS filter transmission curves from \citet{2003AJ....126.1090C} because the curves have already been multiplied by $\lambda$.}.  We computed the residuals ($\delta_{X-Y} = (X-Y)_{\mathrm{obs}} - (X-Y)_{\mathrm{synth}}$) between the observed and synthetic $J-H$, $H-K$, and $K-L'$ colors of each dwarf.  The average residuals are given in Table \ref{residuals} and indicate that no significant errors were introduced in the relative flux density levels of spectra in adjacent orders during the order merging process.

The final step in the reduction process was to absolutely flux calibrate the spectra using the published photometry listed in Table \ref{sample}.  For each spectrum, we computed correction factors given by,

\begin{equation}
C_X = 10^{-0.4m_X} \times\frac{\int \lambda f^{\mathrm{Vega}}_{\lambda}(\lambda)T_X(\lambda)d\lambda}{\int \lambda f^{\mathrm{obs}}_{\lambda}(\lambda)T_X(\lambda)d\lambda} \label{fluxcaleqn}
\end{equation}
\noindent 
where $m_X$ is the magnitude of the object in the $J$, $H$, and $K$ bands and the remaining variables have the same meaning as in Equation 1.  The errors in the correction factors were computed using the errors in the published magnitudes.  Each spectrum was then multiplied by the weighted average of the $J$-, $H$- and $K$-band correction factors.  The average correction factors ranged from 0.9 to 1.05 with a median value of 0.92.

\subsection{IRCS Observations}\label{IRCS Observations}

The IRCS is a facility infrared imager and spectrograph for the 8.2m Subaru telescope.  The imager, which is equipped with a 1024 $\times$ 1024 Alladin 3 InSb array, also contains grisms for low-resolution spectroscopy.  We used the $L$-band grism (2.90$-$4.16 $\mu$m) with slit widths of 0$\farcs$3 and 0$\farcs$6 to achieve spectral resolving powers of $\sim$425 and $\sim$212, respectively.   

A log of the observations, including the UT date of observation, the resolving power $R$, total on-source integration time and A0 V telluric standard star, is given in Table \ref{ircslog}.  The observing strategy was identical to that employed in the SpeX observations.  A series of exposures were taken in pairs with the targets at two different positions separated by 5$\arcsec$ along the 20$\arcsec$-long slit.  A nearby A0 V star was also observed for each science object to correct for absorption due to the Earth's atmosphere and a series of flat field exposures and dark frames was taken at the end of each night.

The spectra were extracted using a modified version of Spextool \citep{2004PASP..116..362C}.  Each pair of target exposures was subtracted and then flat fielded.  The spectra were then extracted and wavelength calibrated using sky emission features.  The telluric correction process is much simpler than in the case of the SpeX observations because at these wavelengths, the continuum of the A0 V standard star is well approximated by a blackbody and the hydrogen lines are relatively weak.  Therefore, we divided the object spectrum by the standard star spectrum to remove the telluric absorption and instrument throughput and then multiplied the result by a Planck function of 9500 K, the effective temperature of A0 V stars \citep{tokunaga00}, to restore the continuum shape of the object.

The IRCS spectra must be absolutely flux calibrated using $L'$-band photometry before they can be combined with the SpeX spectra since the two sets of spectra do not overlap in wavelength.  DENIS 0255$-$4700 lacks a published $L'$ magnitude so we observed it at $L'$ using the SpeX guider camera on 2003 October 03 UT and measured $m_{L'}=10.2\pm0.1$.  After the IRCS spectra are flux calibrated using the $L'$-band photometry they were merged with the SpeX spectra.

\subsection{Combining the Optical and Infrared Spectra}\label{Combining the Optical and Infrared Spectra}

The near-infrared spectra were extended blueward to $\sim$0.6 $\mu$m using published red-optical spectra.  These spectra, which cover from roughly 0.6 to 0.95 $\mu$m and typically have resolutions ranging from 9 to 18 \AA \, ($R$=890$-$444), are from \citet{1991ApJS...77..417K,1999ApJ...519..802K,2000AJ....120..447K}, \citet{1998MNRAS.301.1031T}, \citet{1999AJ....118.2466M}, \citet{2000AJ....119..928F}, \citet{2000AJ....120.1085G}, \citet{2000AJ....119..369R}, \citet{2003ApJ...594..510B} and T. Henry (2004, private communication).  For each dwarf, the wavelengths of the optical spectrum were converted from air to vacuum if necessary.  The optical spectrum was scaled to the same flux level as the flux-calibrated SpeX spectrum and linearly interpolated onto the wavelength grid of the SpeX spectrum.  The two spectra were then combined using a mean.  Since many of the optical spectra have not been corrected for telluric absorption, care was taken to scale the spectra using wavelength ranges with a minimal amount of telluric absorption, namely 0.81$-$0.83 $\mu$m and 0.99$-$1.05 $\mu$m.

\begin{figure}
\includegraphics[width=5.5in]{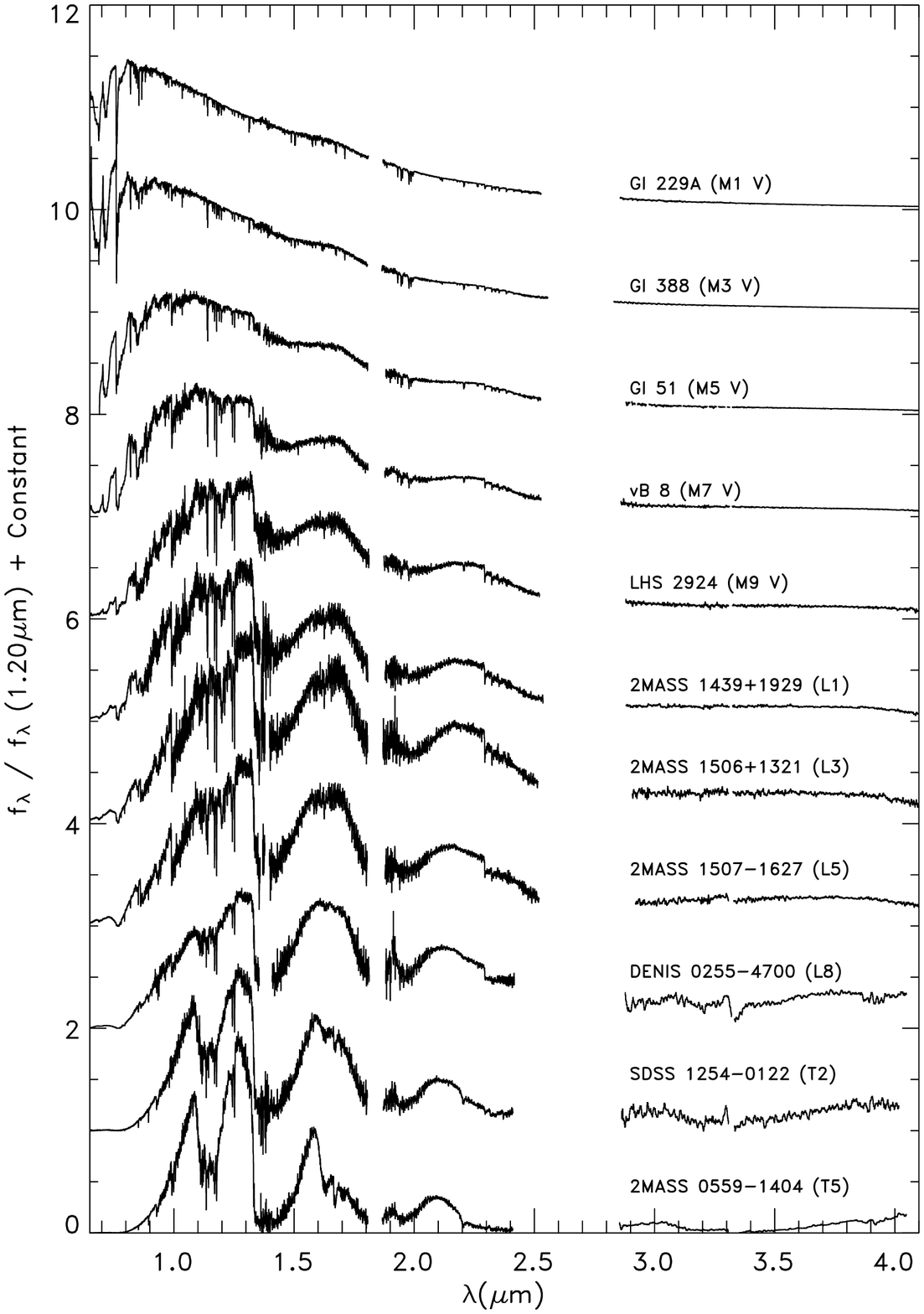}
\caption{\label{fsequence}0.6 to 4.2 $\mu$m sequence of M, L and T dwarfs.  The gaps in the spectra at $\sim$1.85 $\mu$m are due to a break in the wavelength coverage of the SXD mode of SpeX.  The spectra from 2.5 to 2.9 $\mu$m were removed because the atmosphere is opaque at these wavelengths.}
\end{figure}

Figure \ref{fsequence} shows a subset of the flux calibrated 0.6$-$4.1 $\mu$m spectra of the dwarfs listed in Table \ref{sample}.  The S/N of the spectra in the $J$, $H$, and $K$ bands ranges from $\sim$50 in the T dwarfs to $\sim$200 in the M dwarfs.  The S/N in the $L'$ band is typically lower ranging from $\sim$10 in the T dwarfs to $\sim$100 in the M dwarfs.

\section{Analysis} \label{Analysis}

\subsection{Feature Identifications} \label{Feature Identifications}

The identification of atomic and molecular absorption features in the spectra of M, L, and T dwarfs is important because they can be used to constrain the effective temperatures ($T_{\mathrm{eff}}$), surface gravities ($\log g$), and metallicities ([M/H]) of the dwarfs.  In addition, the process allows the identification of opacity sources not currently included in atmospheric models.  There have been numerous infrared spectroscopic studies that identify both atomic and molecular absorption features in the spectra of M, L, and T dwarfs \citep[e.g.,][]{1994MNRAS.267..413J,1995AJ....110.2415A,1996ApJS..104..117L,2000ApJ...533L..45M,2001AJ....121.1710R,2000AJ....120.1100B,2002ApJ...564..421B}.  However, the quality of the identifications ranges from secure to questionable because in some cases, the spectra have low spectral resolution, low S/N, small wavelength coverage and/or sparse spectral type coverage.  Since our data do not suffer from these shortcomings, we have conducted a systematic search for both atomic and molecular features in our M, L, and T dwarf spectra.  We have confined our search to $\lambda \ga 0.95$ $\mu$m since \citet{1991ApJS...77..417K,1995AJ....109..797K,1999ApJ...519..802K} have adequately identified the atomic and molecular features in the spectra of ultracool dwarfs shortward of this wavelength.

We describe the identification of the molecular and atomic features in the following two sub-sections.  Many of the features are weak and could easily be mistaken for noise.  However, we are confident in our identifications because the features can be tracked through the spectral sequence.  Tables \ref{molfeat} lists the identified molecular features while Table \ref{atoms} lists the identified atomic features.  Finally, we avoid the use of the term `lines' since at $R\approx2000$, many of the features, especially the molecular ones, are blends of many individual lines.

\subsubsection{Molecular Features} \label{Molecular Features}

TiO. --- Absorption bands of TiO that arise from the $\gamma$ ($A$ $^3\Phi$ $-$ $X$ $^3\Delta$), $\gamma^{\prime}$ ($B$ $^3\Pi$ $-$ $X$ $^3\Delta$), $\delta$ ($b$ $^1\Pi$ $-$ $a$ $^1\Delta$) and $\epsilon$ ($E$ $^3\Pi$ $-$ $X$ $^3\Delta$) systems are prominent in the red-optical spectra of mid- to late-type M stars \citep[e.g., ][]{1991ApJS...77..417K,1998MNRAS.301.1031T} and early-type L dwarfs \citep{1999ApJ...519..802K}.  The $\phi$ ($b$ $^1\Pi$ $-$ $d$ $^1\Sigma$) system has also been detected in the spectra of giant stars \citep{1998AJ....116.2520J} but never, to our knowledge, in the spectra of dwarf stars.  However, synthetic spectra of M dwarfs computed from the model atmospheres of \citet{2000ApJ...540.1005A} predict a strong 0$-$0 bandhead at $\sim$1.104 $\mu$m.

We have compared the high-resolution (R=800,000) Fourier Transform Spectrometer (FTS) emission spectrum\footnote{ We obtained the spectrum (770629R0.005) using the National Solar Observatory Digital Library Query Tool, http://diglib.nso.edu/nso\_user.html} of TiO used in the analysis of the $\phi$ system by \citet{1980ApJS...42..241G} to our M and L dwarf spectra in order to determine to what extent the 0$-$0 band of the $\phi$ system is present.  We smoothed the TiO spectrum to $R=2000$ and then resampled it onto the wavelength grid of the dwarfs.  We find no evidence of absorption due to the 0$-$0 band in the spectra of the M and L dwarfs.

VO. --- Absorption bands of VO arising from the $B$ $^4\Pi- X$ $^4\Sigma^-$ system are prominent in the red-optical spectra of late-type M dwarfs \citep{1991ApJS...77..417K,1995AJ....109..797K,1998MNRAS.301.1031T}.  The $A$ $^4\Pi- X$ $^4\Sigma^-$ system produces three bands, the 1$-$0 at $\sim$0.96 $\mu$m, the 0$-$0 at $\sim$1.06 $\mu$m, and the 0$-$1 at 1.18 $\mu$m.  The 0$-$0 and 0$-$1 bands and their associated bandheads have been detected in the spectra of M giants \citep[e.g.,][]{1998AJ....116.2520J}; the 1$-$0 band is difficult to detect because of contamination by telluric H$_2$O absorption.  \citet{1993ApJ...402..643K} identified the 0$-$0 band in the spectrum of vB 8 (M7 V) and \citet{1999ApJ...519..802K} claimed to detect this band in the spectrum of GD 165 B (L4).  However, the \citeauthor{1999ApJ...519..802K} spectrum of GD 165 B is missing the wavelengths centered on the 0$-$0 band of VO which casts doubt on the identification.

\begin{figure}
\includegraphics[width=6in]{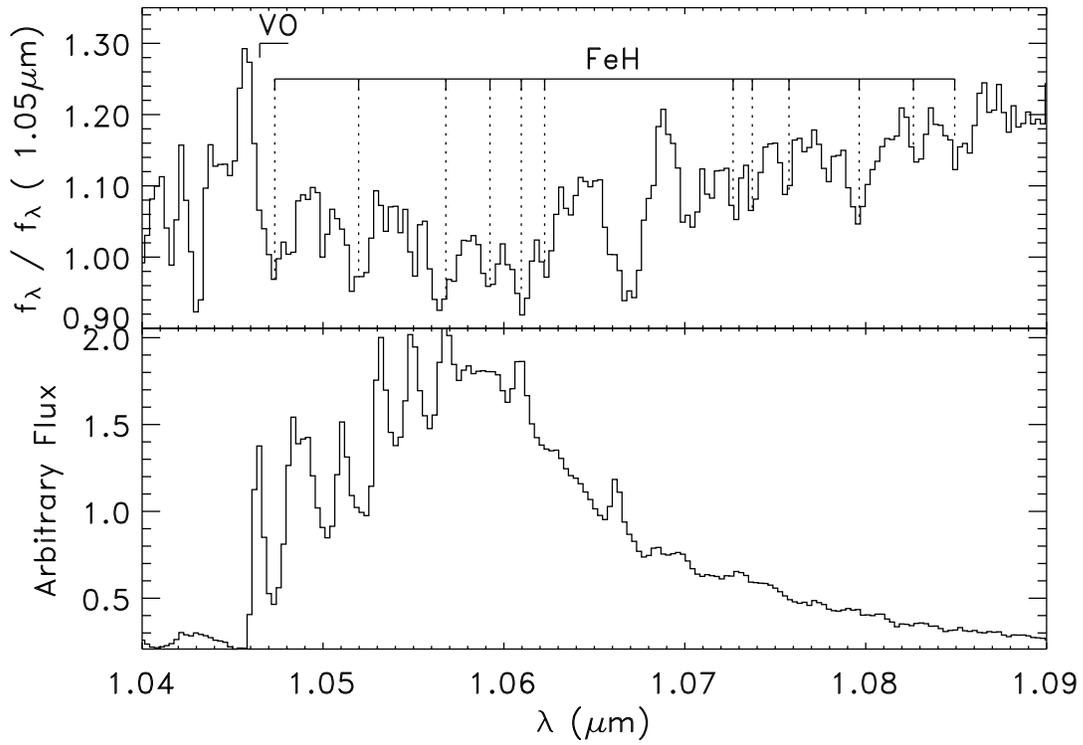}
\caption{\label{VO}The lower panel shows the VO emission spectrum and the upper panel is the spectrum of vB 10 (M8 V).  FeH features previously identified are indicated with dotted lines.  The broad absorption feature in the spectrum is vB 10 is due to VO.}
\end{figure}

We have compared our spectrum of vB 10 (M8 V) with a high-resolution FTS emission spectrum of VO kindly provided S. Davis (2002, private communication) in order to confirm the presence of the 0$-$0 band of VO in the spectra of M and L dwarfs.  The FTS spectrum was smoothed to $R=2000$ and then resampled onto the wavelength grid of vB 10.  Figure \ref{VO} shows the spectrum of vB 10 in the top panel and the emission spectrum of VO in the lower panel.  It is apparent that the broad feature in the vB 10 spectra centered at 1.06 $\mu$m is due to the 0$-$0 band of VO.  However, almost all of the high frequency features seen in the vB 10 spectrum have been ascribed to FeH \citep{2003ApJ...582.1066C}.  It may be that some of the features in this wavelength range identified as FeH are actually a blend of both FeH and VO lines.

\begin{figure}
\includegraphics[width=6in]{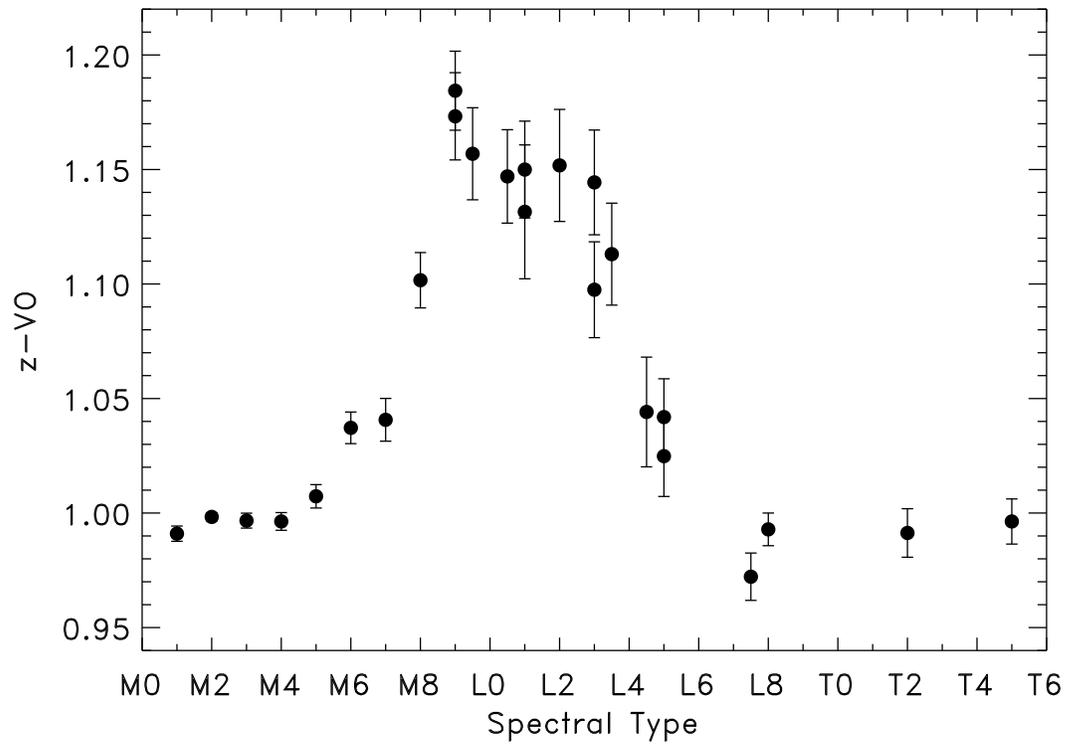}
\caption{\label{VOindex}The $z$-VO index as a function of spectral type.}
\end{figure}

In order to quantify the variation of the strength of this band as a function of spectra type, we followed \citet{1995AJ....109..797K} and defined a spectral index that measures the ratio of the flux density in the band to the expected flux density at the same wavelength based on a linear interpolation between continuum points on either side of the band.  This index is given by,

\begin{equation}
z\mbox{-VO} = \frac{0.4667 f_{1.084} + 0.5334 f_{1.039}}{f_{1.06}},
\end{equation}
\noindent
where $f_{\lambda_{0}}$ is the mean flux density in a 0.008 $\mu$m window centered around $\lambda_0$.  Larger values of $z$-VO imply deeper absorption.  The VO indices computed for the dwarfs in our sample are shown as a function of spectral types in Figure \ref{VOindex}.  The error bars were computed from the uncertainties in the mean flux densities in $f_{1.039}$, $f_{1.06}$, and $f_{1.084}$.  The VO band first appears at a spectral type of $\sim$M6, peaks in strength at $\sim$L0,  and disappears by $\sim$L5.  The strengths of the 1$-$0 and 0$-$0 bands of the $B$ $^4\Pi- X$ $^4\Sigma^-$ system peak at $\sim$M9 \citep{1999ApJ...519..802K}, which lends further credence to the idea that the broad feature at 1.06 $\mu$m is carried by VO.  Given that the \citeauthor{1999ApJ...519..802K} GD 165B spectrum is missing the wavelengths centered on this band, this is the first detection of the 0$-$0 band of the $A$ $^4\Pi- X$ $^4\Sigma^-$ transition of VO in the spectra of L dwarfs.

CrH. --- The appearance of the CrH bandhead, which arises from the 0$-$0 band of the $A$ $^6\Sigma^+$ $-$ $X$ $^6\Sigma^+$ system, at 0.8611 $\mu$m in the spectra of late-type dwarfs is one of the defining characteristics of the L spectral class \citep{1999ApJ...519..802K,1999AJ....118.2466M}.  Although the 0$-$1 bandhead of the same system at $\sim$0.9969 $\mu$m is also often identified in the spectra of late-type dwarfs, \citet{2003ApJ...582.1066C} have shown that in $R=2000$ spectra many of the absorption features near this wavelength can be attributed to the 0$-$0 band of the $F$ $^4\Delta- X$ $^4\Delta$ system of FeH.  As a result, the extent to which the 0$-$1 band is present in the spectra of late-type dwarfs remains uncertain until higher resolution spectra can be obtained.

\citet{2002ApJ...577..986B} have computed new line lists and opacities for 12 bands of the $A$ $^6\Sigma^+$ $-$ $X$ $^6\Sigma^+$ system of CrH.  In addition to the 0$-$0 and 0$-$1 bands, this system also has bands in the near-infrared at $\sim$1.18 $\mu$m (0$-$2) and $\sim$1.4 $\mu$m (0$-$3).  The 0$-$3 band will be difficult to identify because it falls within the strong water absorption band, both telluric and intrinsic to the dwarfs, centered at the same wavelength.  In order to determine the extent to which the 0$-$2 band is present in the spectra of L dwarfs, we have compared the CrH cross-section spectrum of \citet{2002ApJ...577..986B} with the $J$-band spectrum of 2MASS 1507$-$1627 (L5).  We chose an L5 dwarf to maximize our chances of identifying any CrH features since the strength of the 0$-$0 band at 0.8611 $\mu$m peaks at this spectral type \citep{1999ApJ...519..802K}.  Nevertheless we find no evidence of CrH absorption features in the $J$-band spectrum of 2MASS 1507$-$1627.  \citet{2003ApJ...594..510B} noted that the weakness of the CrH band at 0.9969 $\mu$m in L dwarf spectra is probably due to the low CrH/FeH equilibrium abundance ratio of $\sim$10$^{-3}$ for 1800 $\leq T_{\mathrm{eff}} \leq$ 2500 K \citep{1999ApJ...519..793L} since the cross-sections of FeH and CrH are approximately equal at 0.99 $\mu$m.  A similar effect may occur in the $J$ band since the 0$-$1 band of FeH at $\sim$1.19 $\mu$m is strong in the spectra of L dwarfs \citep{2003ApJ...582.1066C,2003ApJ...596..561M} and the cross-sections of CrH and FeH are also approximately equal at these wavelengths \citep{2003ApJ...594..651D}.

CH$_4$. --- The appearance of overtone and combination bands of CH$_4$ in the near-infrared spectra of late-type dwarfs is the defining characteristic of the T spectral class \citep{2002ApJ...564..421B}.  For the most part, published spectra of T dwarfs have low spectral resolution so the absorption bands appear relatively smooth.  At a resolving power of $R\approx1200$, it should be possible to identify individual CH$_4$ absorption features.  To this end, we have compared an FTS emission spectrum of CH$_4$ \citep{2003JQSRT..82..279N} with our spectrum of 2MASS 0559$-$1404 (T5.0) to search for any common features.  The CH$_4$ spectrum, which was obtained at a temperature of 1000 K,  covers from 1.56 to 5.0 $\mu$m at a resolving power of $R \approx 160,000$.  We smoothed the CH$_4$ spectrum to $R=1200$ and resampled it onto the wavelength grid of 2MASS 0559$-$1404.

\begin{figure}
\includegraphics[width=6in]{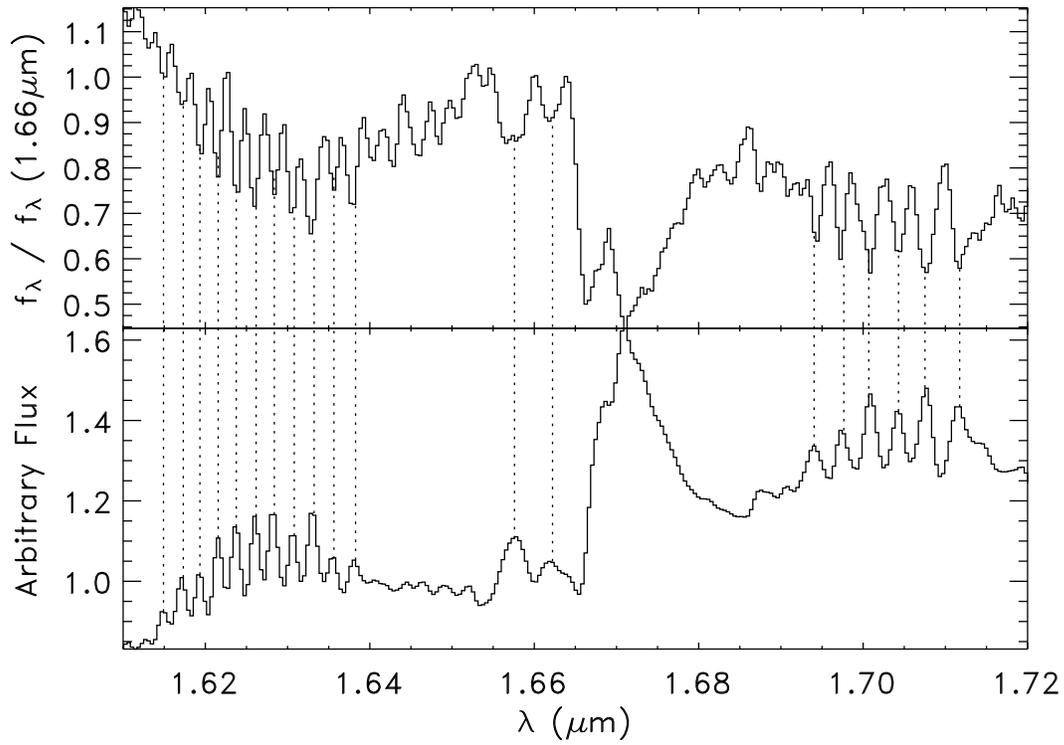}
\caption{\label{ch4comph} The lower panel shows the CH$_4$ emission spectrum and the upper panel shows the spectrum of 2MASS 0559$-$1404 (T5) centered on the 2$\nu_3$ band.  The positions of the 19 features common to both spectra are indicated with dotted lines.}
\end{figure}

Figure \ref{ch4comph} shows the spectra of CH$_4$ and 2MASS 0559$-$1404 (T5) centered on the 2$\nu_3$ band ($\sim$1.6 $\mu$m).  Nineteen features common to the two spectra are shown as dotted lines and are listed in Table \ref{ch4features}.  Although the spectra of the T dwarfs exhibit a series of roughly equidistant absorption features from 1.615 to 1.65 $\mu$m, the same series of emission features in the CH$_4$ spectrum is truncated at $\sim$1.635 $\mu$m.  In addition, the strong absorption feature in the T dwarf spectra at 1.665 $\mu$m is missing in the CH$_4$ emission spectrum.  The lack of emission of features at these wavelengths is most likely caused by self-absorption due to room temperature CH$_4$ in the apparatus \citep{2003JQSRT..82..279N}.  These features can also be seen throughout the T dwarf sequence \citep{2003ApJ...596..561M}.

\begin{figure}
\includegraphics[width=6in]{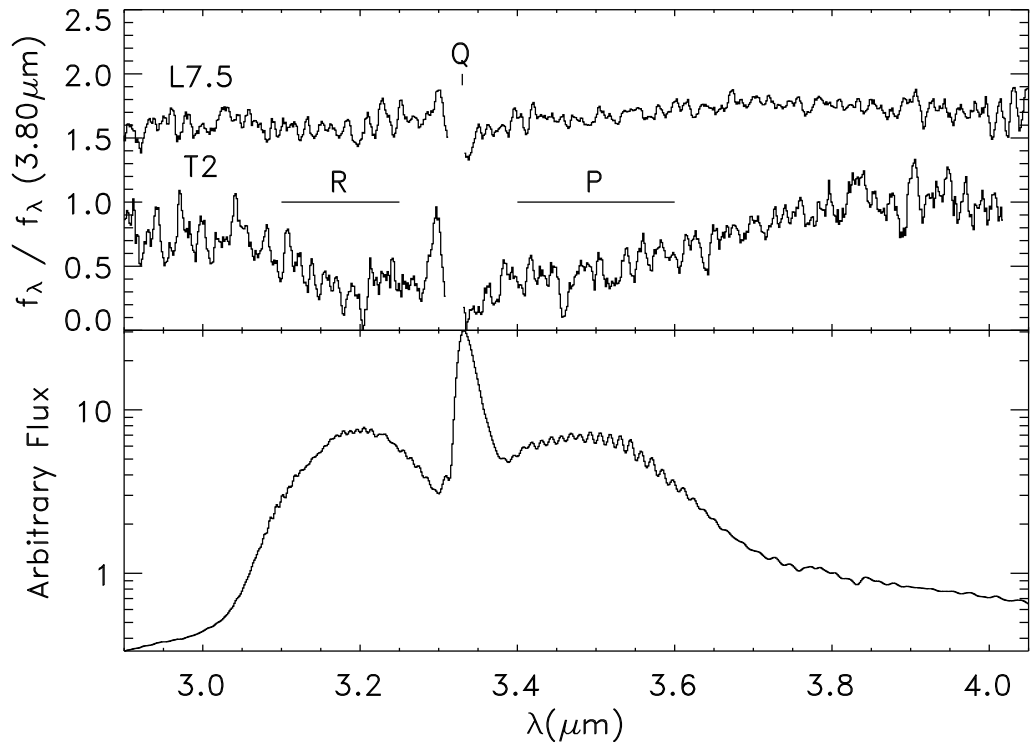}
\caption{\label{ch4compl}The lower panel shows the CH$_4$ emission spectrum and the upper panel shows the spectra of 2MASS 0825$+$2115 (L7.5) and SDSS 1254$-$0122 (T2).  Note the CH$_4$ spectrum is plotted on a logarithmic scale.  The positions of the P, Q, and R branches are indicated.}
\end{figure}

The Q-branch of the $\nu_3$ fundamental band of CH$_4$ at 3.33 $\mu$m has also been detected in the spectra of two late-type L dwarfs, 2MASS 1507$-$1627 (L5) and 2MASS 0825+2115 (L7.5), by \citet{2000ApJ...541L..75N}.  Although the detection is secure, the spectra were of fairly low S/N.  Figure \ref{ch4compl} shows the emission spectrum of CH$_4$ in the lower panel and the spectra of 2MASS 0825$+$2115 and SDSS 1254$-$0122 (T2) in the upper panels, respectively.  The dwarf spectra ($R\approx300$) were obtained using the IRCS on Subaru (see \S\ref{IRCS Observations}) and have a higher S/N than the \citet{2000ApJ...541L..75N} spectra.  The Q-branch is clearly seen in both dwarf spectra while the broad P and R branches can be seen in the spectrum of SDSS 1254$-$0122.

\subsubsection{Atomic Features} \label{Atomic Features}

\begin{figure}
\includegraphics[width=5.5in]{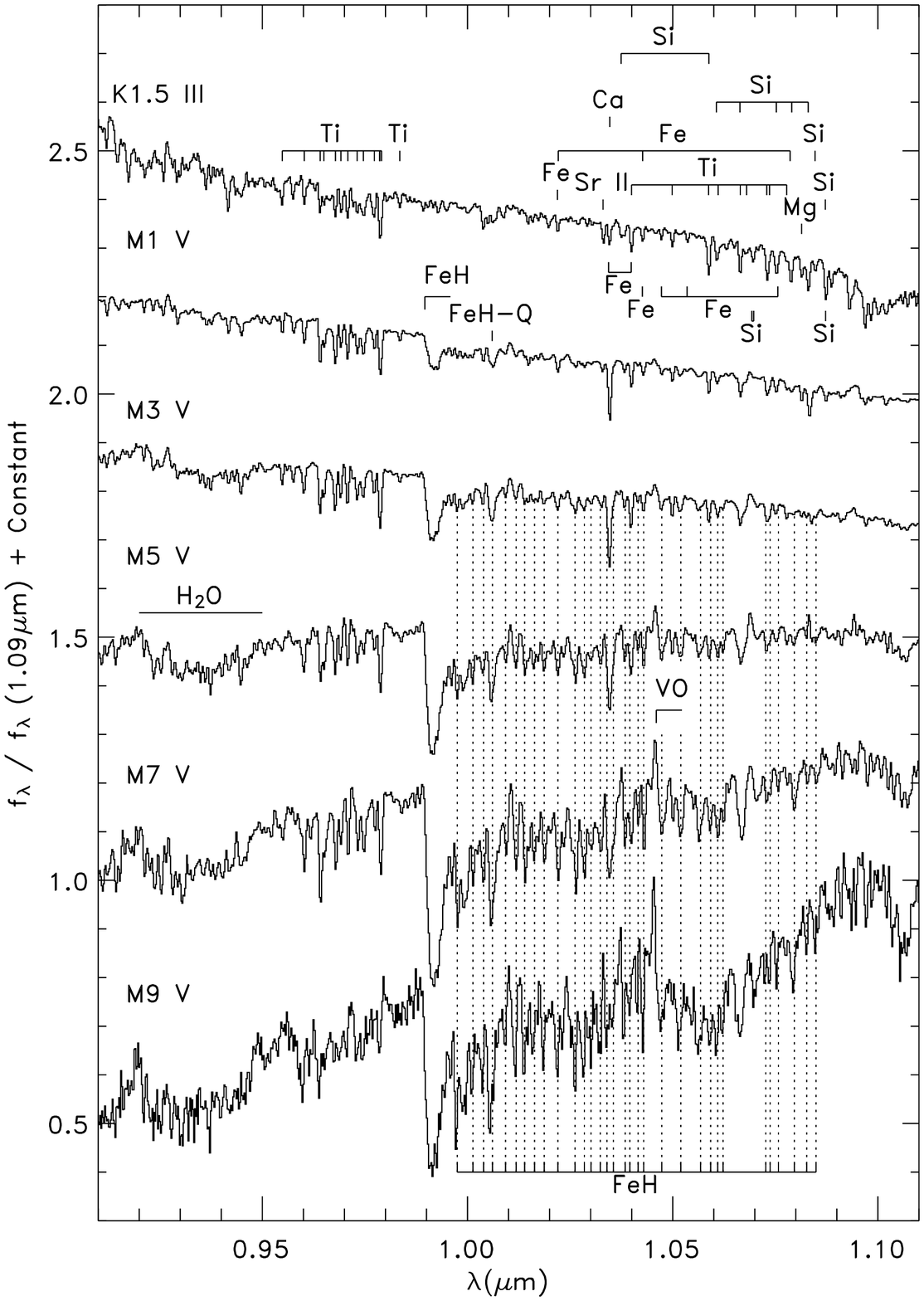}
\caption{\label{Mzseq} $z$-band spectra of Arcturus (K1.5 III), Gl 229A (M1 V), Gl 388 (M3 V), Gl 51 (M5 V) vB 8 (M7 V) and LHS 2924 (M9 V).  The most prominent molecular and atomic features are indicated.} 
\end{figure}

\begin{figure} 
\includegraphics[width=5.5in]{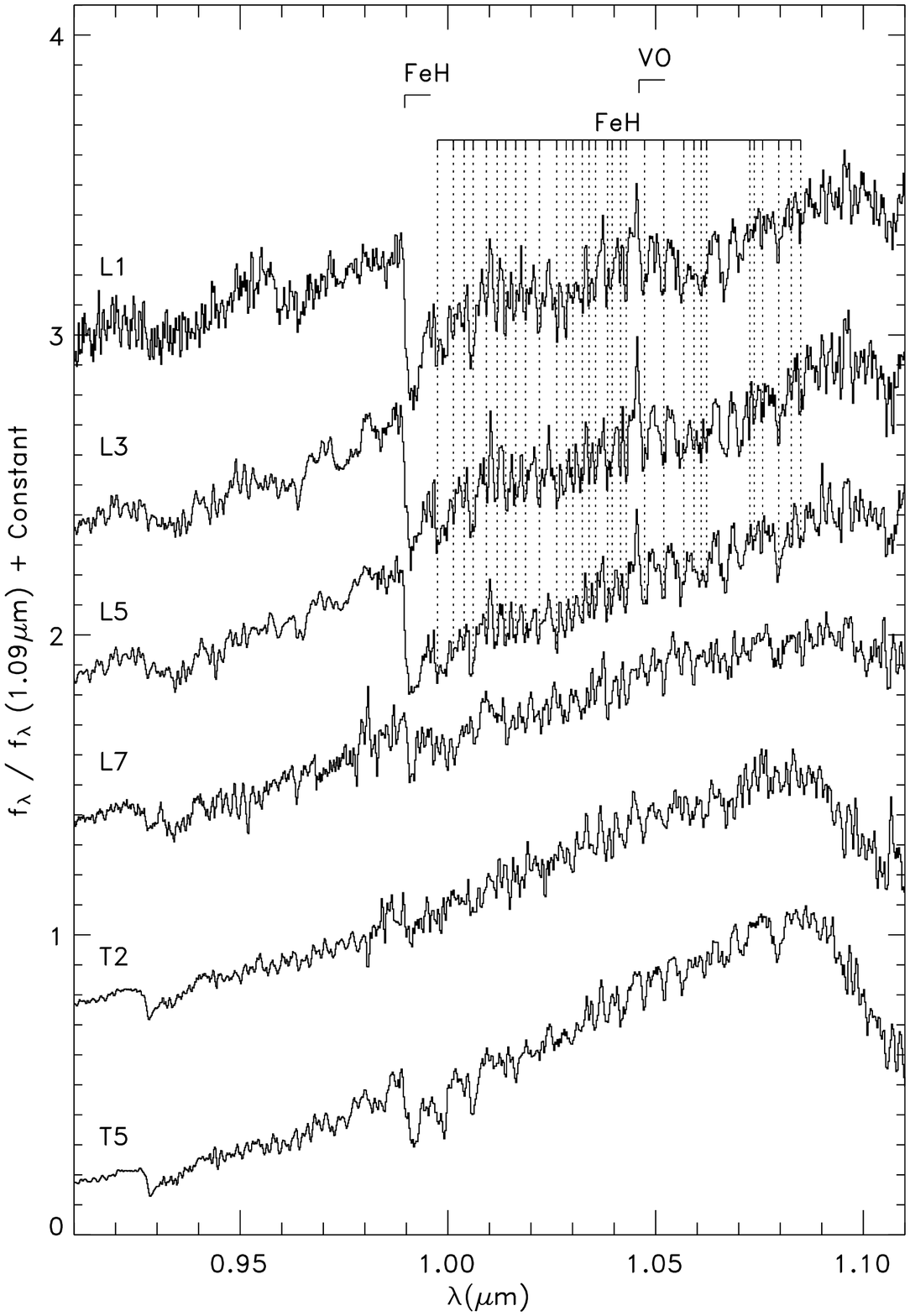}
\caption{\label{LTzseq} $z$-band spectra of 2MASS1439$+$1929 (L1), 2MASS 1506$+$1321 (L3), 2MASS 1507$-$1627 (L5), DENIS 0255$-$4700 (L8), SDSS 1254$-$0122 (T2), and 2MASS 0559$-$1404 (T5).  The most prominent molecular and atomic features are indicated.}
\end{figure}

\begin{figure}
\includegraphics[width=5.5in]{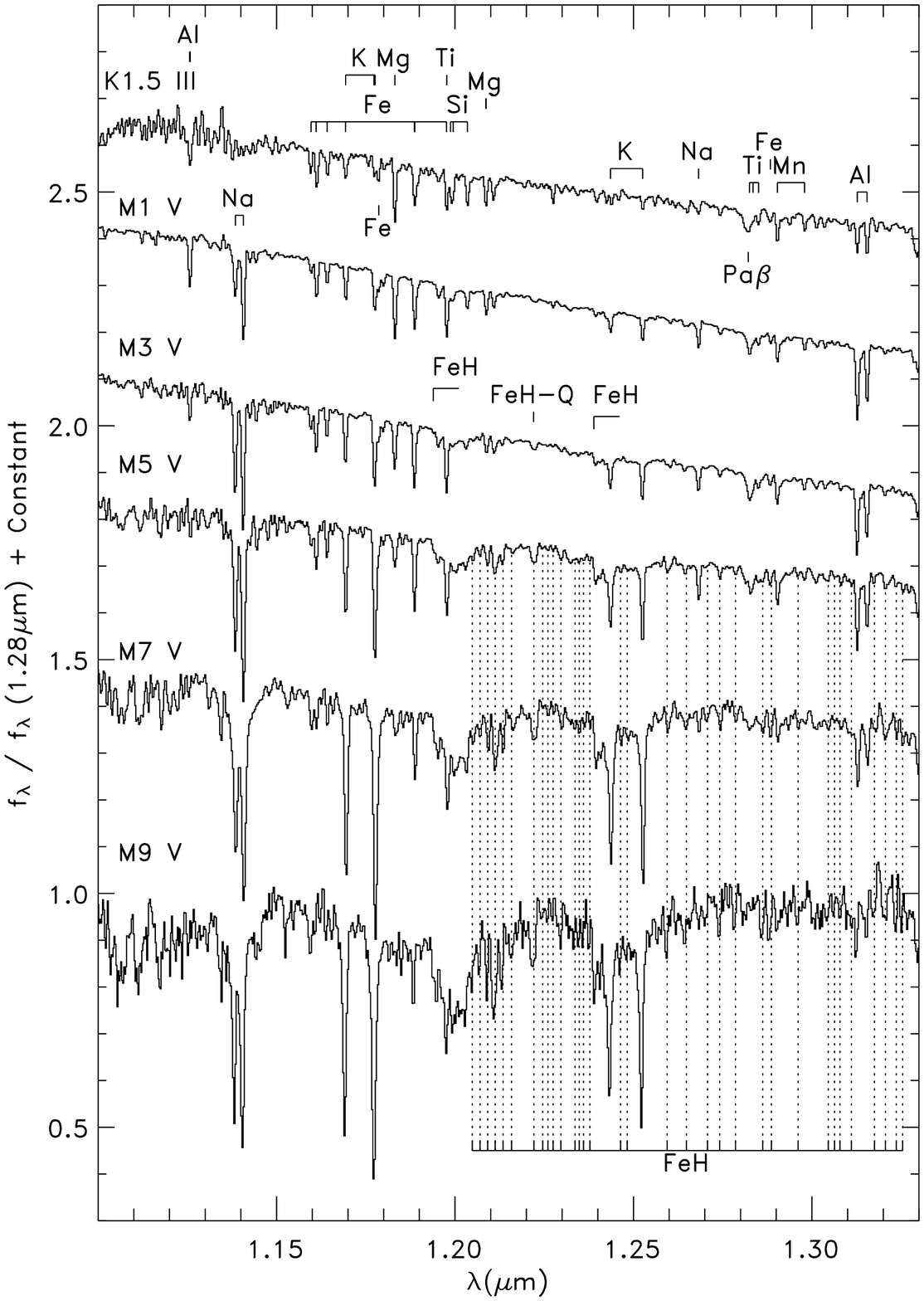}
\caption{\label{MJseq} $J$-band spectra of Arcturus (K1.5 III), Gl 229A (M1 V), Gl 388 (M3 V), Gl 51 (M5 V) vB 8 (M7 V) and LHS 2924 (M9 V).  The most prominent molecular and atomic features are indicated.} 
\end{figure}

\begin{figure}
\includegraphics[width=5.5in]{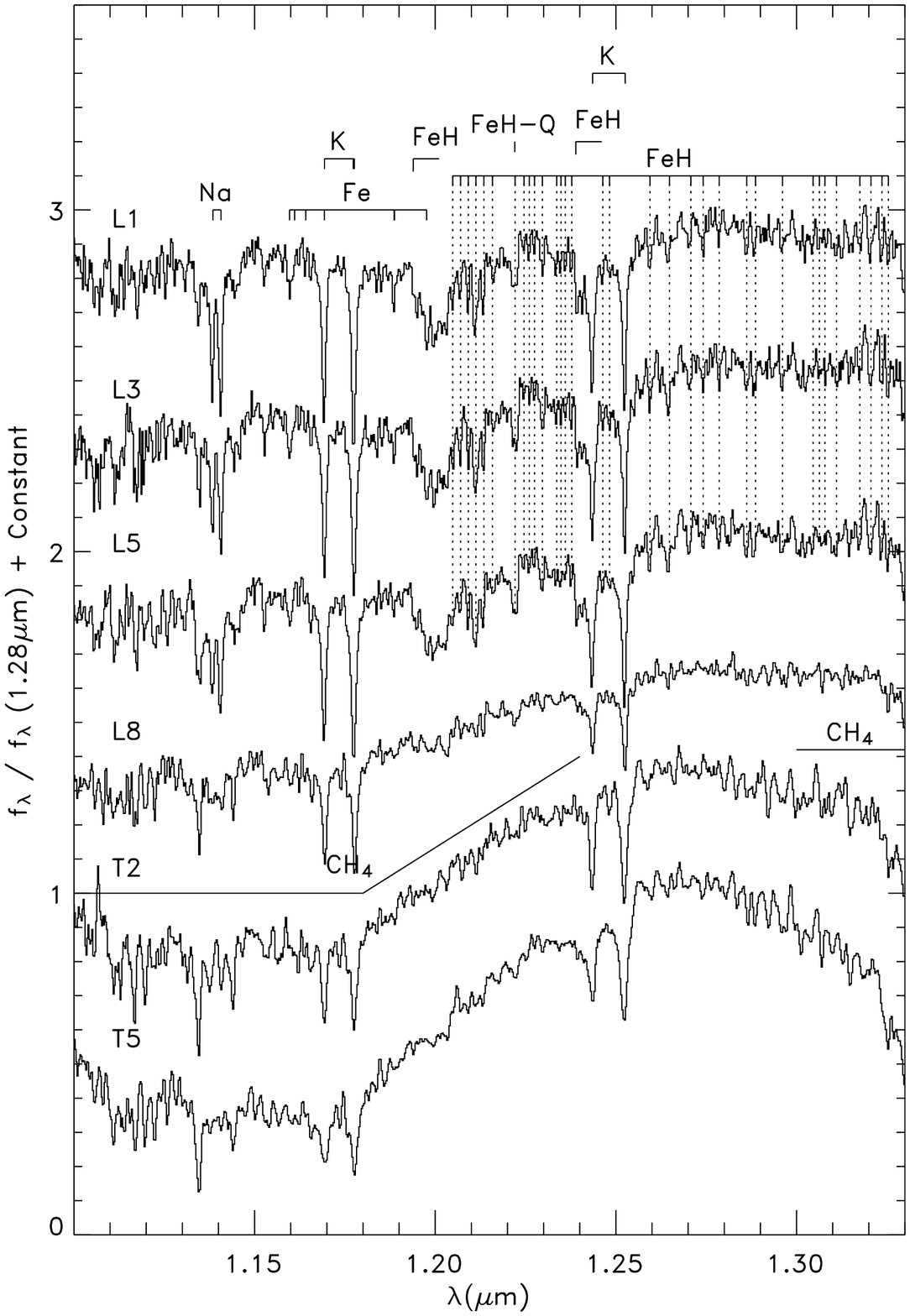}
\caption{\label{LTJseq} $J$-band spectra of 2MASS1439$+$1929 (L1), 2MASS 1506$+$1321 (L3), 2MASS 1507$-$1627 (L5), DENIS 0255$-$4700 (L8), SDSS 1254$-$0122 (T2), and 2MASS 0559$-$1404 (T5).  The most prominent molecular and atomic features are indicated.}
\end{figure}

\begin{figure}
\includegraphics[width=5.5in]{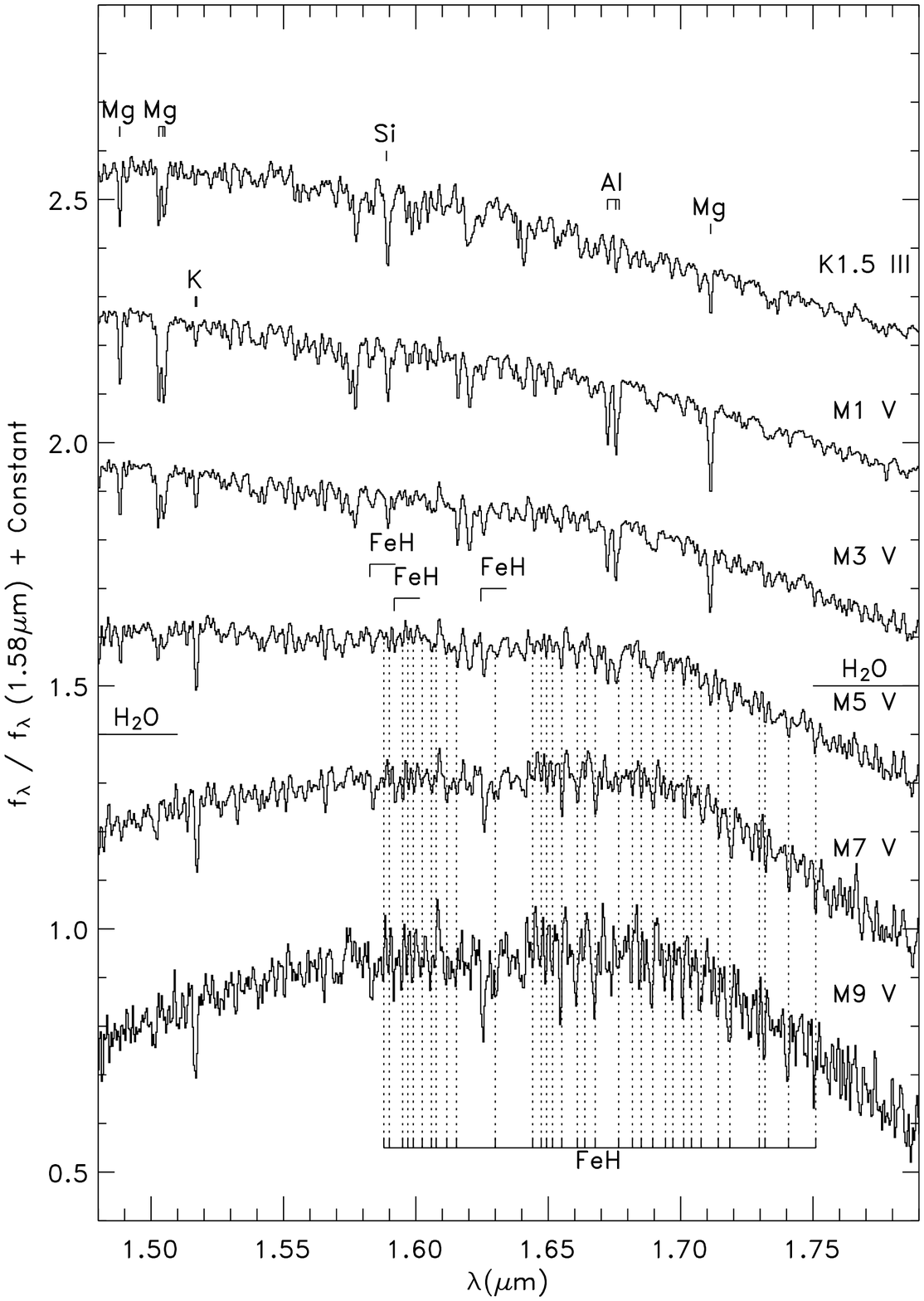}
\caption{\label{MHseq} $H$-band spectra of Arcturus (K1.5 III), Gl 229A (M1 V), Gl 388 (M3 V), Gl 51 (M5 V) vB 8 (M7 V) and LHS 2924 (M9 V).  The most prominent molecular and atomic features are indicated.} 
\end{figure}

\begin{figure}
\includegraphics[width=5.5in]{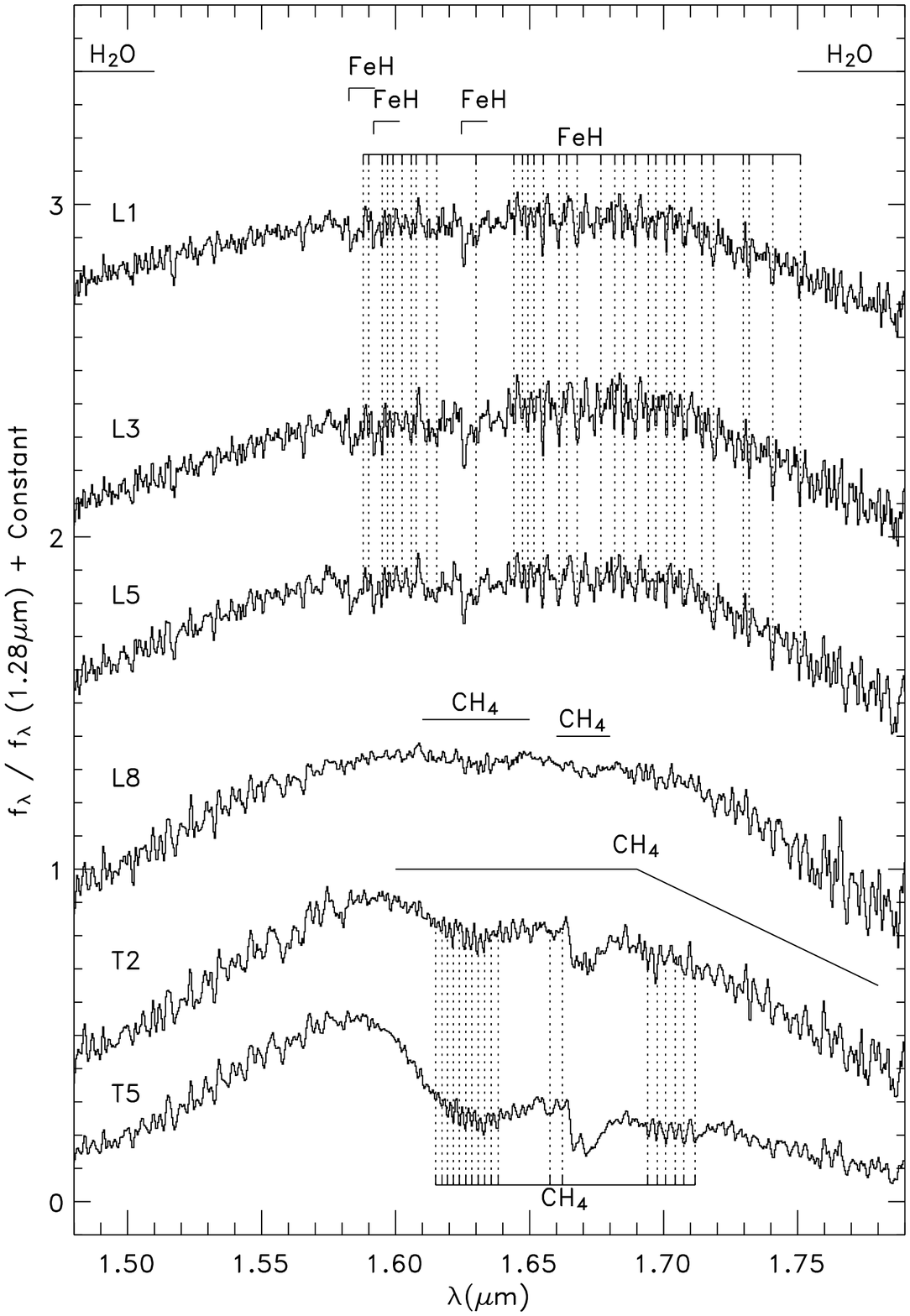}
\caption{\label{LTHseq} $H$-band spectra of 2MASS1439$+$1929 (L1), 2MASS 1506$+$1321 (L3), 2MASS 1507$-$1627 (L5), DENIS 0255$-$4700 (L8), SDSS 1254$-$0122 (T2), and 2MASS 0559$-$1404 (T5).  The most prominent molecular and atomic features are indicated.}
\end{figure}

\begin{figure}
\includegraphics[width=5.5in]{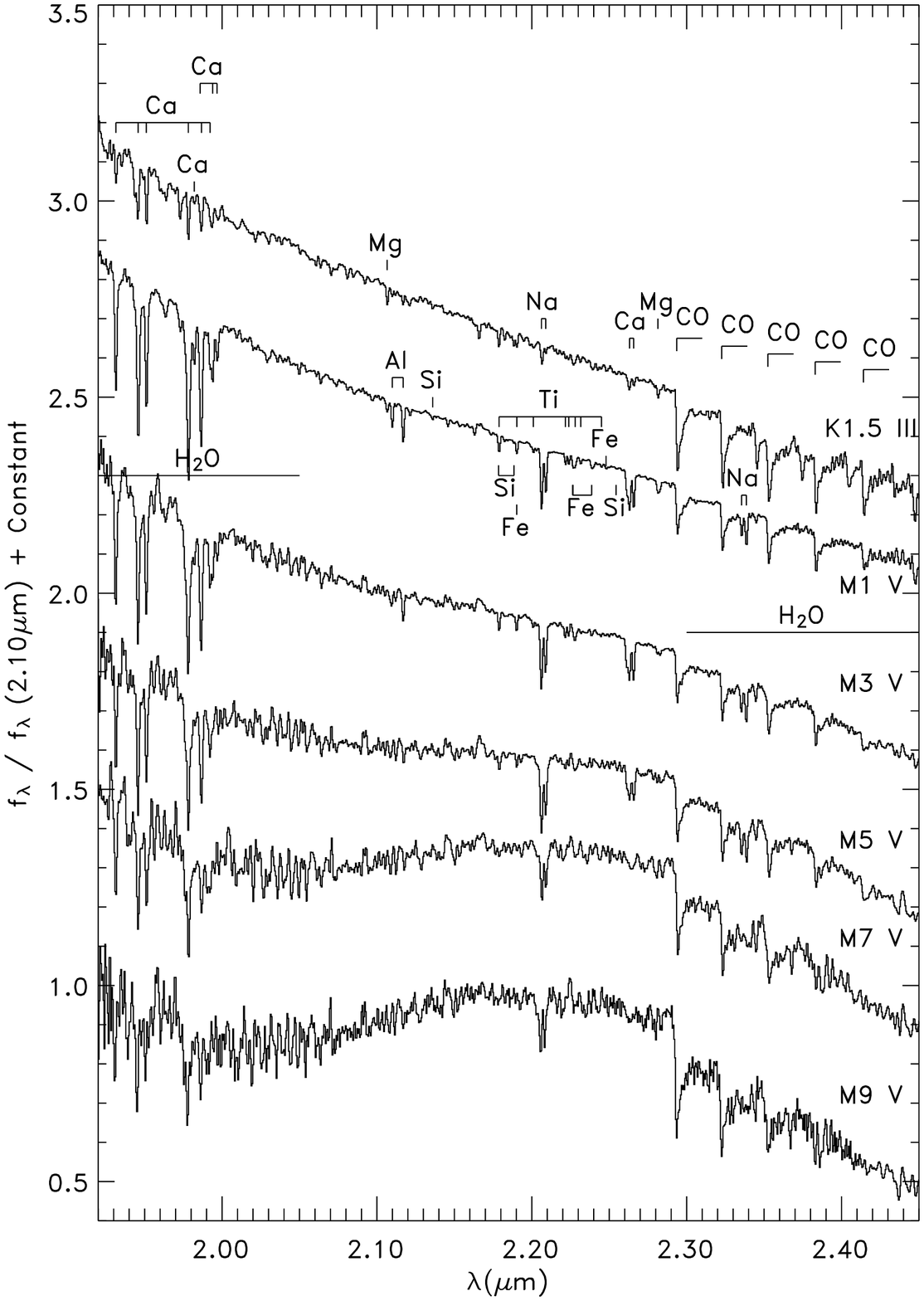}
\caption{\label{MKseq} $K$-band spectra of Arcturus (K1.5 III), Gl 229A (M1 V), Gl 388 (M3 V), Gl 51 (M5 V) vB 8 (M7 V) and LHS 2924 (M9 V).  The most prominent molecular and atomic features are indicated.} 
\end{figure}

\begin{figure}
\includegraphics[width=5.5in]{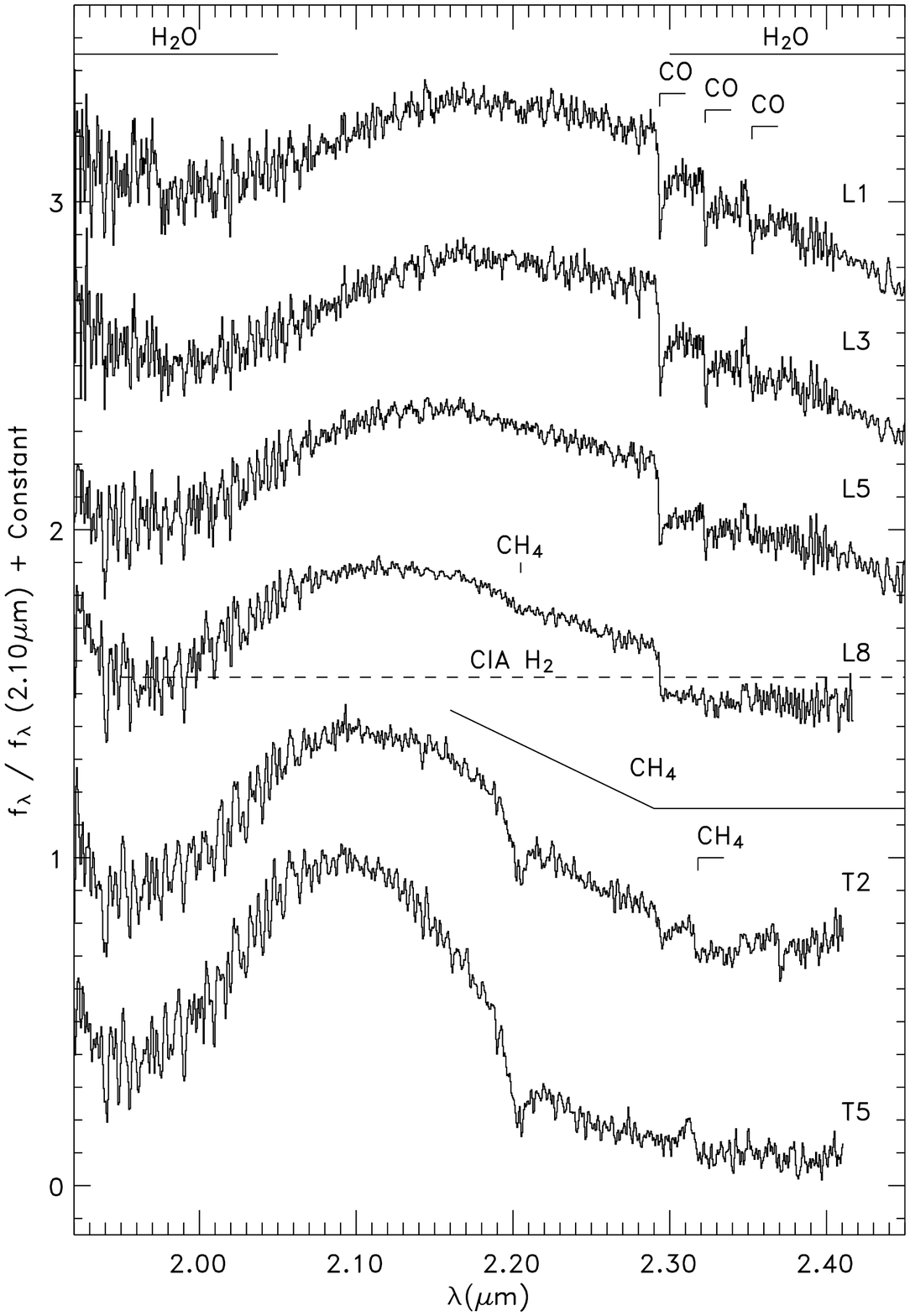}
\caption{\label{LTKseq} $K$-band spectra of 2MASS1439$+$1929 (L1), 2MASS 1506$+$1321 (L3), 2MASS 1507$-$1627 (L5), DENIS 0255$-$4700 (L8), SDSS 1254$-$0122 (T2), and 2MASS 0559$-$1404 (T5).  The most prominent molecular and atomic features are indicated.}
\end{figure}

In order to identify the atomic features in the spectra of the M and L dwarfs, we used the high-resolution ($R$=100,000) infrared spectral atlas of Arcturus \citep{1995PASP..107.1042H} and a spectrum of Arcturus obtained with SpeX.  Arcturus' spectral type is K1.5 III, cool enough that most of the atomic features present in its spectrum will also be present in the spectra of early-type M dwarfs.  We first identified absorption features present in both the SpeX Arcturus spectrum and the spectrum of Gl 229A (M1 V), and then searched the Arcturus atlas for the feature identifications.  The absorption features, which include neutral lines of Ti, Al, Si, Ca, Fe, Mg, K, and Na, are listed in Table \ref{atoms} and shown in Figures \ref{Mzseq} - \ref{LTKseq}.

\citet{2000ApJ...533L..45M} reported a detection of the Rb I line at 1.3234 $\mu$m in the spectra of L dwarfs as well as the Cs I line at 1.359 $\mu$m in the spectrum of 2MASS 1507$-$1627 (L5).  We see no evidence of these lines nor of any other lines of Rb and Cs \citep{1959ApJ...130..683F} in our spectra.  There is an absorption feature at 1.3234 $\mu$m coincident with the position of a Rb line but this feature can be seen in the spectrum of vB 10 (M8 V) and therefore is most likely due to H$_2$O.

\subsection{Spectral Morphology} \label{Spectral Morphology}

Having identified both atomic and molecular absorption features in the spectra of the M, L, and T dwarfs, we now give a detailed description of the spectral morphology of the spectra in the $z$-, $J$-, $H$-, $K$-, and $L$-bands.  Figures \ref{Mzseq} - \ref{LTKseq} show a sequence of the M, L and T dwarfs in each of the $z$-, $J$-, $H$-, $K$- and $L$-bands along with the identified features.

\subsubsection{$z$ Band} \label{z Band}

The $z$-band spectra of the early M dwarfs contain mainly atomic absorption features of Ti, Fe, Ca, Si, and Mg.  Most prominent is a series of 12 Ti features centered at 0.97 $\mu$m arising from the $a$ $^5$F $-$ $z$ $^5$F$^{\circ}$ multiplet.  These features are clearly discernible down to a spectral type of M9 and disappear at $\sim$L1.  The band head of the Wing-Ford band of FeH (0$-$0 band of the $F$ $^4\Delta$ $-$ $X$ $^4\Delta$ system) at 0.9969 $\mu$m,  along with the FeH Q-branch feature at 1.006 $\mu$m, already appear quite strong at a spectral type of M1.  At later spectral types, the Wing-Ford band replaces the atomic features as the dominant carrier of absorption features in this wavelength range.  We have identified 32 FeH features in this wavelength range \citep{2003ApJ...582.1066C}.  These features, along with the 0$-$0 bandhead and Q-branch feature, weaken considerably by a spectral type of $\sim$L7.  The bandhead is also present in the spectrum of the T5 dwarf 2MASS 0559$-$1404 (see \citet{2002ApJ...571L.151B}).  The 0$-$0 band of the $A$ $^4\Pi- X$ $^4\Sigma^-$ system of VO at $\sim$1.06 $\mu$m appears at a spectral type of $\sim$M5, peaks in strength at M9, and then disappears by L5 (see \S\ref{Molecular Features}).  Finally the overall spectral shape becomes progressively redder with spectral type.

\subsubsection{$J$ Band} \label{J Band}

The $J$ band contains the most prominent atomic features in the spectra of M and L dwarfs: the Na doublet at 1.14 $\mu$m, and the two K doublets at 1.175 and 1.25 $\mu$m.  Early M dwarf spectra also exhibit absorption features due to Fe, Mg, Ti, Si, H, and Mn but most of these features weaken in the mid- to late-type M dwarfs. However, the Fe feature at 1.189189 $\mu$m persists to a spectral type of $\sim$L5.  The 0$-$1 bandhead of FeH ($F$ $^4\Delta$ $-$ $X$ $^4\Delta$ system) at 1.1939 $\mu$m and the associated Q-branch feature at 1.22210 $\mu$m, along with the 1$-$2 bandhead at 1.2389 $\mu$m, first appear at a spectral type of M3 and strengthen through the M sequence.  Beginning at a spectral type of $\sim$M5, 33 additional FeH absorption features can also be seen in the range 1.2$-$1.3 $\mu$m \citep{2003ApJ...582.1066C}.  At a spectral type of $\sim$L1, the only remaining atomic features are the Na, and K lines and possibly a few weak Fe features at $\sim$1.155 $\mu$m.  The Na feature at 1.14 $\mu$m and the FeH features are not apparent at a spectral type of $\sim$L7 but the K lines persist through the T sequence \citep{2003ApJ...596..561M}.  As with the $z$ band, the overall spectral shape of the $J$ band becomes progressively redder with spectral type.  Finally, CH$_4$ bands from 1.15$-$1.25 $\mu$m and CH$_4$ and H$_2$O bands longward of 1.28 $\mu$m cause the $J$-band spectra of the T dwarfs to become peaked at $\sim$1.27 $\mu$m.

\subsubsection{$H$ Band} \label{H band}

The $H$ band is the most difficult wavelength range in which to identify features in the spectra of early M dwarfs because it contains many relatively weak absorption features.  Only a few doublets and triplets of Mg, Si, Al, and K are clearly evident.  \citet{1998ApJ...508..397M} identify eight second overtone ($\Delta \nu=3$) bandheads of $^{12}$CO in $R=3000$ spectra of K and M dwarfs.  However, we cannot unambiguously identify any of these bands.  H$_2$O bands on either side of the $H$-band also begin to appear at a spectral type of $\sim$M4 and strengthen through the M, L and T sequence.

A new band of FeH, tentatively identified as the $E$ $^4\Pi$ $-$ $A$ $^4\Pi$ system by \citet{2001ApJ...559..424W}, appears at $\sim$M5 and becomes the dominant carrier of absorption features from 1.59 to 1.75 $\mu$m  \citep{2003ApJ...582.1066C}.  At the same time, the atomic absorption features of Mg, Si, and Al weaken and disappear, except for the K I line at 1.516 $\mu$m which persists to $\sim$L5.  Finally, the appearance of the 2$\nu_3$ and 2$\nu_2 + \nu_3$ bands of CH$_4$ at $\sim$1.67 $\mu$m signals the transitions to the T spectral class although weak CH$_4$ absorption features are also found in the spectra of the late L dwarfs \citep{2003ApJ...596..561M} (see also \S\ref{DENIS 0255-4700 and the L/T Transition}).  Both bands strengthen through the T sequence \citep{2003ApJ...596..561M}.

\subsubsection{$K$ Band} \label{K band}

The $K$-band spectra of early M dwarfs also exhibit atomic features of Ca, Mg, Al, Si, Na, Ti, and Fe.  The most prominent features are the series of Ca lines at 1.95 $\mu$m, the Na doublets at 2.21 and 2.34 $\mu$m, and the Ca doublet at 2.26 $\mu$m.  These features all weaken with increasing spectral type; the 2.26 $\mu$m Ca doublet disappears at $\sim$M7, while the 1.95 $\mu$m Ca lines, as well as the Na lines, disappear at $\sim$L0.  Also prominent are the series of first overtone band heads ($\Delta \nu=+2$) of $^{12}$CO extending redward from 2.29 $\mu$m.  Beginning with the 6$-$4 bandhead at 2.41414 $\mu$m and moving to lower transitions, the bandheads disappear through the M and L dwarfs until only the 2$-$0 is present at a spectral type of T2.  H$_2$O absorption bands also appear on either side of the $K$-band at a spectral type of $\sim$ M4 and strengthen through the M, L, and T sequence.  CH$_4$ absorption at 2.2 $\mu$m first appears in the late-type L dwarfs (see also \S\ref{DENIS 0255-4700 and the L/T Transition}) and continues to strengthen through the T sequence suppressing the $K$-band flux.  The $K$-band flux in T dwarfs is further supressed by collision-induced H$_2$ absorption \citep{2002A&A...390..779B}.  Although this opacity source exhibits no bandheads or distinct spectral features, its presence can be inferred from model atmospheres (M. Marley 2004, private communication) .

\subsubsection{$L$ Band} \label{L Band}

The $L$-band spectra of M and L dwarfs are dominated by molecular absorption features due mainly to H$_2$O.  However, there are also absorption features due to the $\Delta \nu=+1$ bands of OH.  Features from both the 1$-$0 and 2$-$1 bands are seen \citep{2002AJ....124.3393W}.  Like the $z$ and $J$ bands, the overall slope of the spectra becomes progressively redder with spectral type until the spectra are nearly flat at a spectral type of L1.  At a spectral type of $\sim$L5, the $\nu_3$ fundamental band of CH$_4$ appears in the spectra \citep{2000ApJ...541L..75N}.  At first, only the Q-branch is present in the spectra but the P- and R-branches are apparent at T2, and the band becomes saturated by a spectral type of T5.

\subsection{Spectral Indices} \label{Spectral Indices}

\begin{figure}
\includegraphics[width=5.5in]{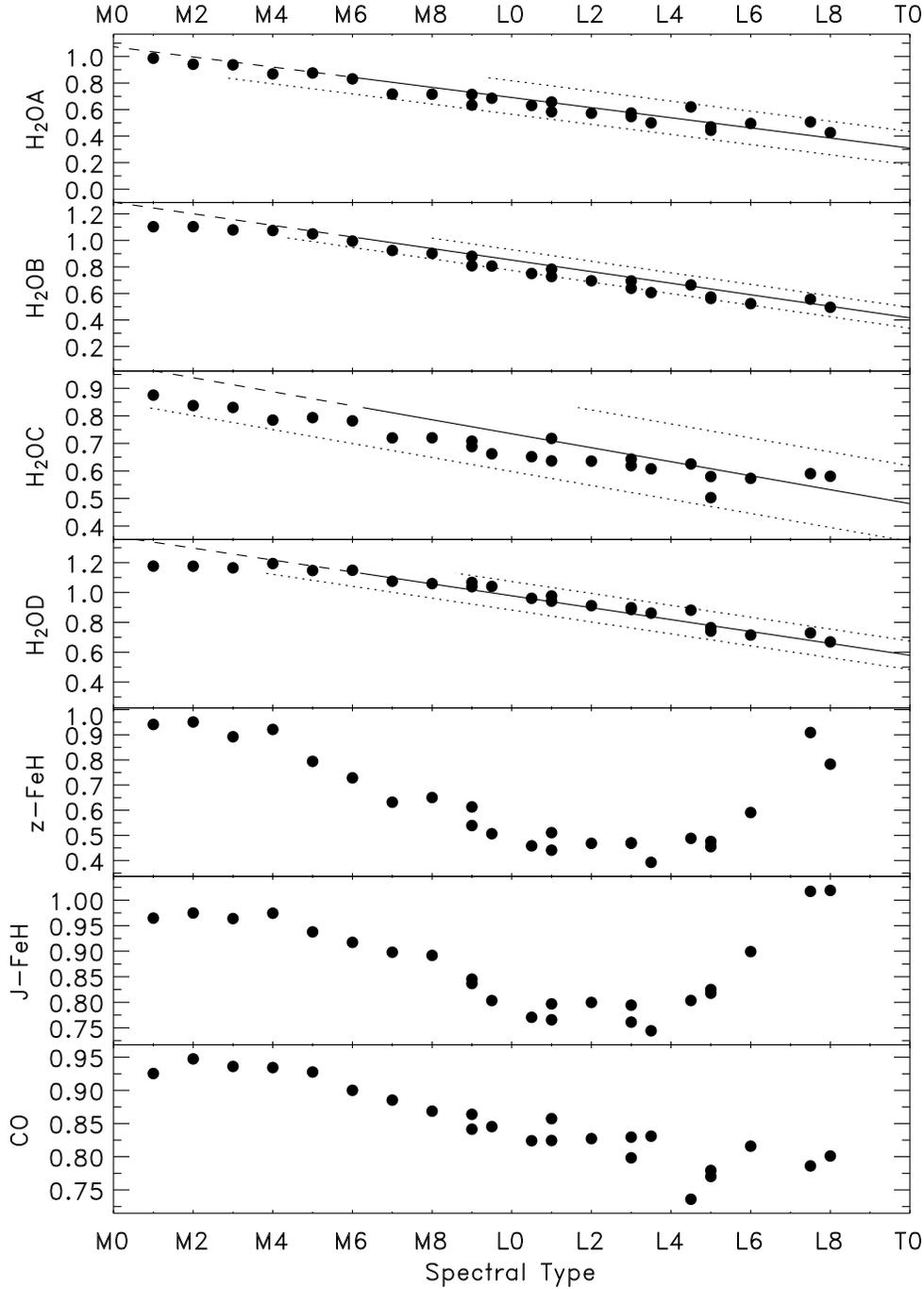}
\caption{\label{mclean}The values of the spectral indices of \citet{2003ApJ...596..561M} for the M and L dwarfs in our sample as a function of spectral type.  Also shown for the H$_2$O indices are the best fitting linear relations derived by \citet{2003ApJ...596..561M} from their data.  The solid lines show the spectral type range over which the lines were derived (M6 to T8) while the dashed lines show the extension of the solid lines to earlier spectral types.  The dotted lines denote the 3 $\sigma$ uncertainies on the relations.}
\end{figure}

As described in the previous section, the overall infrared spectral morphology of the M, L, and T dwarfs, as well as the strengths of many of the absorption features including H$_2$O, FeH, and CO, exhibit a smooth variation with spectral type.  To quantify these variations, numerous spectral indices, ratios of fluxes in particular wavelength intervals chosen to measure the changes in spectral morphology or depths of specific absorption features, have been defined \citep[e.g.,][]{1999AJ....117.1010T,2002ApJ...564..421B,2002ApJ...564..466G,2001ApJ...552L.147T,2001AJ....121.1710R,2003ApJ...596..561M}.  In particular, \citet{2003ApJ...596..561M} have defined a suite of indices for late-type M through T dwarfs (M6 to T8) using spectra with resolving powers similar to that of our spectra ($R$=2000).  The H$_2$OA, H$_2$OB, H$_2$OC, and H$_2$OD indices measure the strength of the 1.35 $\mu$m, 1.4 $\mu$m, 1.7 $\mu$m and 2.0 $\mu$m H$_2$O water bands, respectively, while the $z$-FeH, J-FeH, and CO indices measure the strengths of the 0$-$0 band of FeH at 0.99 $\mu$m, the 0$-$1 FeH bandhead at 1.17 $\mu$m, and the $^{12}$CO 2$-$0 bandhead at 2.29 $\mu$m, respectively. We have computed these indices for the dwarfs in our sample in order to determine how well they characterize the spectra of dwarfs with spectral types earlier than M6 V.  Figure \ref{mclean} shows the values of the indices as a function of spectral type for the M and L dwarfs in our sample; a smaller value implies stronger absorption.

The H$_2$OA and H$_2$OC indices show a good correlation with spectral type throughout the entire M and L sequence.  However the continuum flux window for the H$_2$OA index (1.313 $\mu$m) is centered on the Al I doublet at 1.3130514 $\mu$m (see \S\ref{Equivalent Widths of Atomic Features}) and as a result, cannot be used for M dwarfs with spectral types earlier than $\sim$M8.  The H$_2$OB and H$_2$OD indices are roughly constant from M0 to M4 at which point they decrease steadily through the M, L, T sequence as show by \citet{2003ApJ...596..561M}.  Also shown in Figure \ref{mclean} are the best fitting straight lines (solid/dashed) and $\pm$3 $\sigma$ errors (dotted) derived by \citet{2003ApJ...596..561M}.  All of our data are within the 3 $\sigma$ errors but our H$_2$OC indices lie systematically below the line defined by McLean et al.  It is unclear what causes the systematic differences for the H$_2$OC index.

The two FeH indices show similar behavior.  The indices are roughly constant from M0 to $\sim$M4 at which point they decrease steadily until a spectral type of $\sim$L0.  The indices are roughly constant from L0 to L5, although there is significant scatter.  The FeH bands disappear by the end of the L sequence (the indices approach unity) but the 0$-$0 band reappears in the mid-type T dwarfs \citep{2002ApJ...571L.151B,2003ApJ...596..561M}. Finally the CO index values, like the H$_2$OB, H$_2$D, $J$-FeH, and $z$-FeH indices, are roughly constant from M0 to $\sim$M4 and then decrease through the entire L sequence.  The CO index, like the FeH indices, shows significant scatter in the L sequence.  Overall, the \citeauthor{2003ApJ...596..561M} indices can be used to constrain spectral types only for those dwarfs later than M4.

\subsection{Equivalent Widths of Atomic Features} \label{Equivalent Widths of Atomic Features}

We computed the equivalent widths (EWs) of the strongest atomic features in the spectra of the dwarfs.  The EW and the variance of the EW, $\sigma_{EW}^2$, are given by,

\begin{eqnarray}
EW & = & \sum_{i=1}^{n} \left [ 1 - \frac{f(\lambda_i)}{f_c(\lambda_i)} \right ] \Delta\lambda_i \\
\sigma_{EW}^2 & = & \sum_{i=1}^{n} \Delta\lambda_i^2 \left [\frac{\sigma^2(\lambda_i)}{f_c^2(\lambda_i)} + \frac{f^2(\lambda_i)}{f_c^4(\lambda_i)} \sigma_c^2(\lambda_i) \right ]
\end{eqnarray}
\noindent
where $f(\lambda_i)$ and $f_c(\lambda_i)$ are the observed and estimated continuum flux densities, $\sigma(\lambda_i)$ and $\sigma_c(\lambda_i)$ are the errors in the observed and estimated continuum flux densities, $n$ is the number of wavelength intervals across the feature, and $\Delta \lambda_i$ is the width of wavelength interval $i$.  We estimated $f_c(\lambda_i)$, $\sigma(\lambda_i)$, and $\sigma_c(\lambda_i)$ following the procedure of \citet{1992ApJS...83..147S}.  A linear unweighted polynomial is fitted to the values of the continuum in the regions on either side of the feature in question to determine $f_c(\lambda_i)$.  Assuming the fit to be good, $\sigma(\lambda_i)$ is determined \textit{a posteriori} by setting it equal to the standard deviation, $s$, of the flux density values around $f_c(\lambda_i)$ in the fitting regions.  The errors in the continuum, $\sigma_c(\lambda_i)$, are then computed by multiplying the covariance matrix produced by the least-squares fit by $s$ and then using the standard error propagation formula, including the coefficient covariances \citep[see Eqn. A13 of ][]{1992ApJS...83..147S}.

\begin{figure}
\includegraphics[width=5.5in]{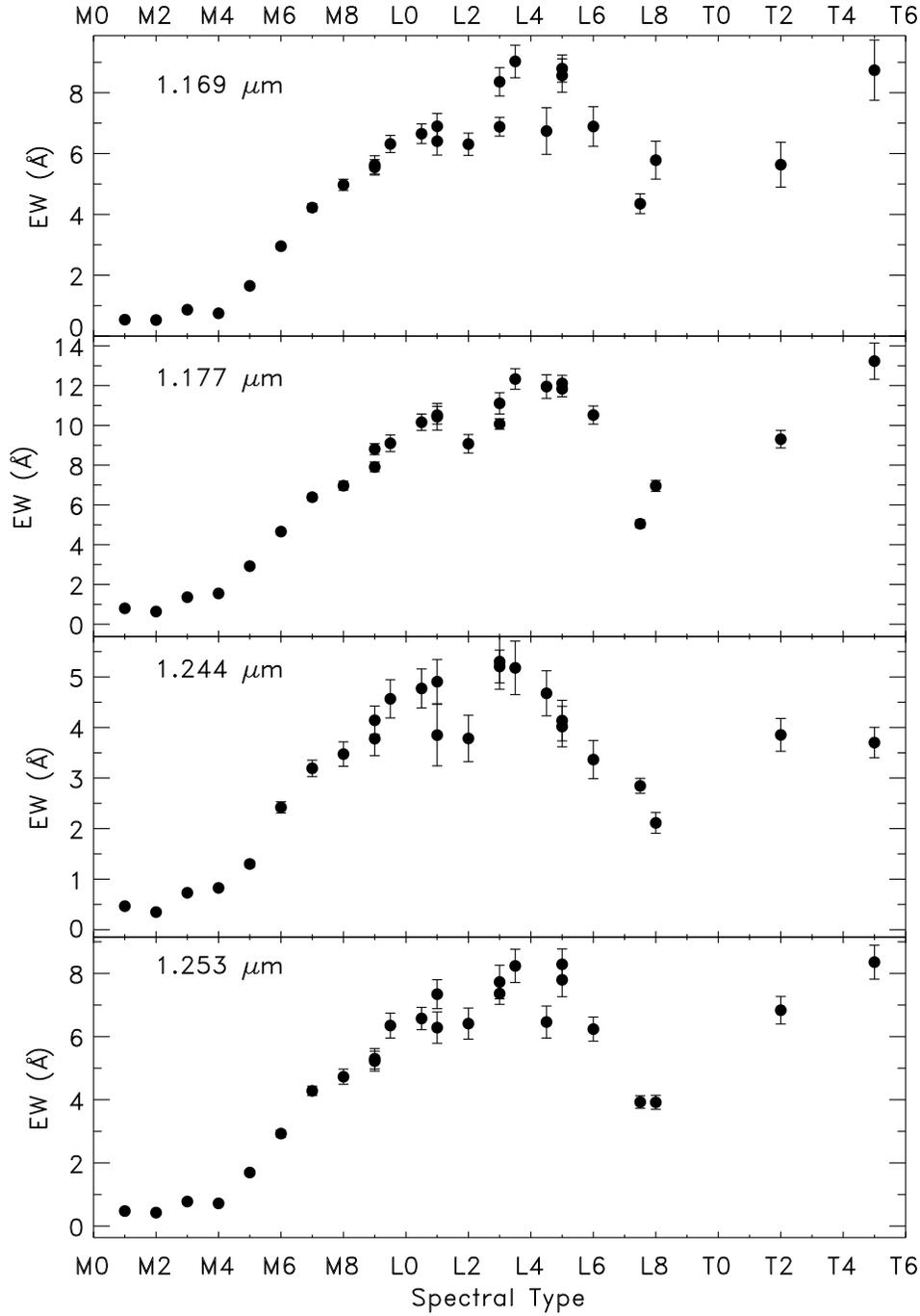}
\caption{\label{KIewp} The EWs of the K I lines in the spectra of the dwarfs in our sample as a function of spectral type.}
\end{figure}

The strongest atomic features in the spectra of late-type dwarfs are the two K I doublets located at $\sim$1.175 $\mu$m and $\sim$1.245 $\mu$m. Their EWs are tabulated in Table \ref{KIew} and are shown as a function of spectral type in Figure \ref{KIewp}.  Although there is some scatter, the EWs of all four lines show a similar behavior with spectral type:  they are roughly constant until a spectral type of $\sim$M4 at which point they increase markedly and reach a maximum at a spectral type of $\sim$L3.  The EWs then decrease through the late L dwarfs but appear to increase again in the T dwarfs.  The strengthening of the K I lines in the early T dwarfs was originally noted by \citet{2002ApJ...564..421B} who ascribed the behavior to the competing effects of decreasing effective temperature and loss of cloud opacity.

\begin{figure}
\includegraphics[width=5.5in]{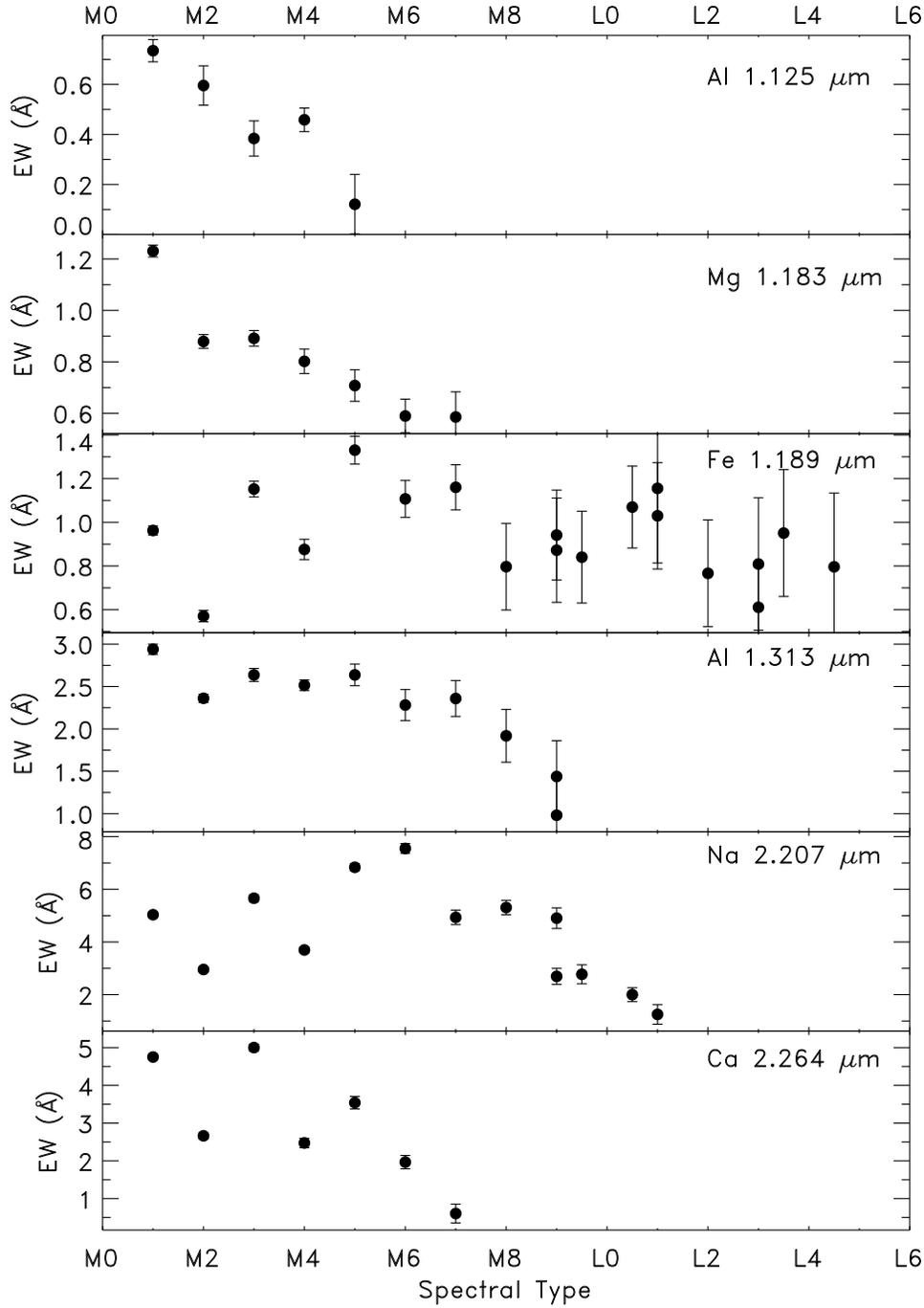}
\caption{\label{metalsewp} The EWs of prominent lines of  Al, Fe, Mg, Ca, and Na in the spectra of the dwarfs in our sample as a function of spectral type.}
\end{figure}

We have also computed the EWs of a number of weaker atomic features including the Na doublet at 2.207 $\mu$m, and a number of features from refractory elements, including Al, Mg, Fe, and Ca.  Except for the Mg line at 1.183 $\mu$m, all of the features are a blend of a number of lines.  Their EWs are tabulated in Table \ref{metalsewt} and are shown as a function of spectral type in Figure \ref{metalsewp}.  There is significant scatter but all the lines weaken with increasing spectral type except for the Fe I line at 1.189 $\mu$m which shows almost no correlation with spectral type.  Additionally, all of the lines disappear by a spectral type of $\sim$L1 except for the Fe I line which persists to $\sim$L5.  The refractory element lines probably weaken in the late-type M dwarfs due to the formation of condensates in the atmospheres of the dwarfs \citep{2002ApJ...577..974L} but the effects of decreasing effective temperature may also play a role.  Two stars, Gl 411 (M2 V) and Gl 213 (M4 V), have low Fe, Al, Na, and Ca EWs relative to the trend established by the other M dwarfs suggesting they may have low metallicities. The classification of Gl 213 as an old disk-halo star by \citet{1992ApJS...82..351L} suggests it may have a low metallicity while the metallicity of Gl 411 has been measured to be [Fe/H] = $-$0.2 \citep{1978ApJ...226..923M}.  We will explore the variations in the EWs of these features more fully in a subsequent paper.

\section{Individual Objects of Interest} \label{Individual Objects of Interest}

\subsection{2MASS 2224$-$0158} \label{2MASS 2224-0158}

\begin{figure}
\includegraphics[width=6in]{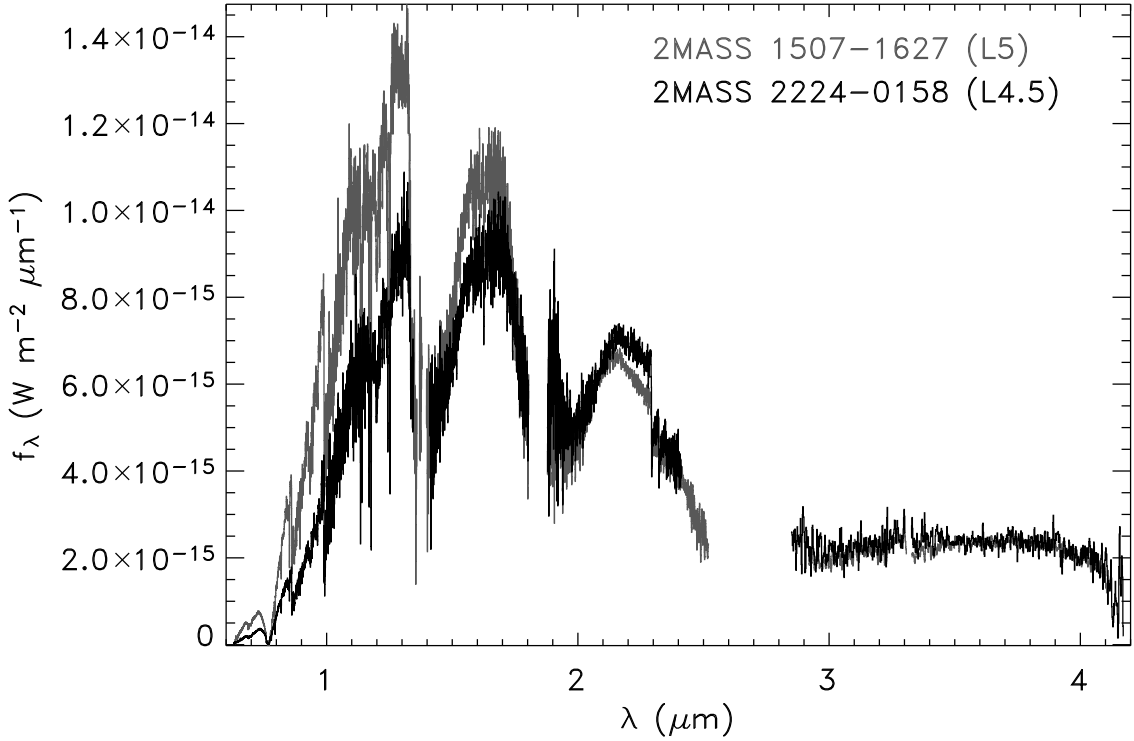}
\caption{\label{2mass2224}0.6 to 2.4 $\mu$m spectra of 2MASS 2224$-$0158 (L4.5) and 2MASS 1507$-$1627 (L5).  The spectrum of 2MASS 1507$-$1627 has been scaled to appear as if the object were at the distance of 2MASS 2224$-$0158.  The $J$- and $H$-band peaks of 2MASS 2224$-$0158 are suppressed relative to that of 2MASS 1507$-$1627 and the $^{12}$CO bandhead in the spectrum of 2MASS 2224$-$0158 is considerably deeper than in the spectrum of 2MASS 1507$-$1627.}
\end{figure}

2MASS 2224$+$0158 is classified as L4.5 based on its red-optical spectrum \citep{2000AJ....120..447K}.  The 0.6$-$4.1 $\mu$m spectrum of 2MASS 2224$-$0158 is shown in Figure \ref{2mass2224} along with the spectrum of 2MASS 1507$-$1627, a normal L5 dwarf.  The two objects have the same spectral type within the errors ($\pm$0.5 subclass).  The 2MASS 1507$-$1627 spectrum was scaled by the square of the ratio of the distances of the two objects to adjust its flux to the level that would be observed if it were at the distance of 2MASS 2224$-$0158.  The spectrum of 2MASS 2224$-$0158 is depressed in the $J$ and $H$ bands relative to that of 2MASS 1507$-$1627, which results in a very red color $J-K = 2.051$.  The $^{12}$CO 2$-$0 and 3$-$1 bandheads at $\sim$2.3 $\mu$m in 2MASS 2224$-$0158 are also considerably deeper than in the spectrum of 2MASS 1507$-$1627.

Interstellar extinction is probably not the cause of the extremely red color since 2MASS 2224$-$0158 lies at a distance of $\sim$11 pc.  An examination of the blue POSS I plates also reveals no obvious extinction in the field of 2MASS 2224$-$0158.  Therefore the extremely red color of 2MASS 2224$-$0158 appears to be intrinsic to the object and may result from an unusually thick cloud deck, since dwarfs become progressively redder with an increasing column abundance of dust.  The CO $\Delta \nu=+2$ bandheads in the spectra of M giant stars are deeper than in the spectra of M dwarfs of the same spectral type \citep{1986ApJS...62..501K}.  If this trend were to continue to cooler temperatures, the deep CO bands may indicate that 2MASS 2224$-$0158 has a low surface gravity.

\subsection{DENIS 0255$-$4700 and the L/T Transition} \label{DENIS 0255-4700 and the L/T Transition}

There has been much interest in objects with spectral types at the transition between the L and T dwarfs \citep{2001ApJ...561L.115M,2001ApJ...561L.119N,2002ApJ...566..435S,2002ApJ...564..466G,2002ApJ...564..421B,2003ApJ...585L.151T}.  This interest stems not only from a desire to determine the exact spectral morphological changes that signal the transition to the T spectral class but also to understand why dwarfs at the L/T boundary (L7 to T3) have roughly the same effective temperature \citep{2004AJ....127.3516G}.

\begin{figure}
\includegraphics[width=5.5in]{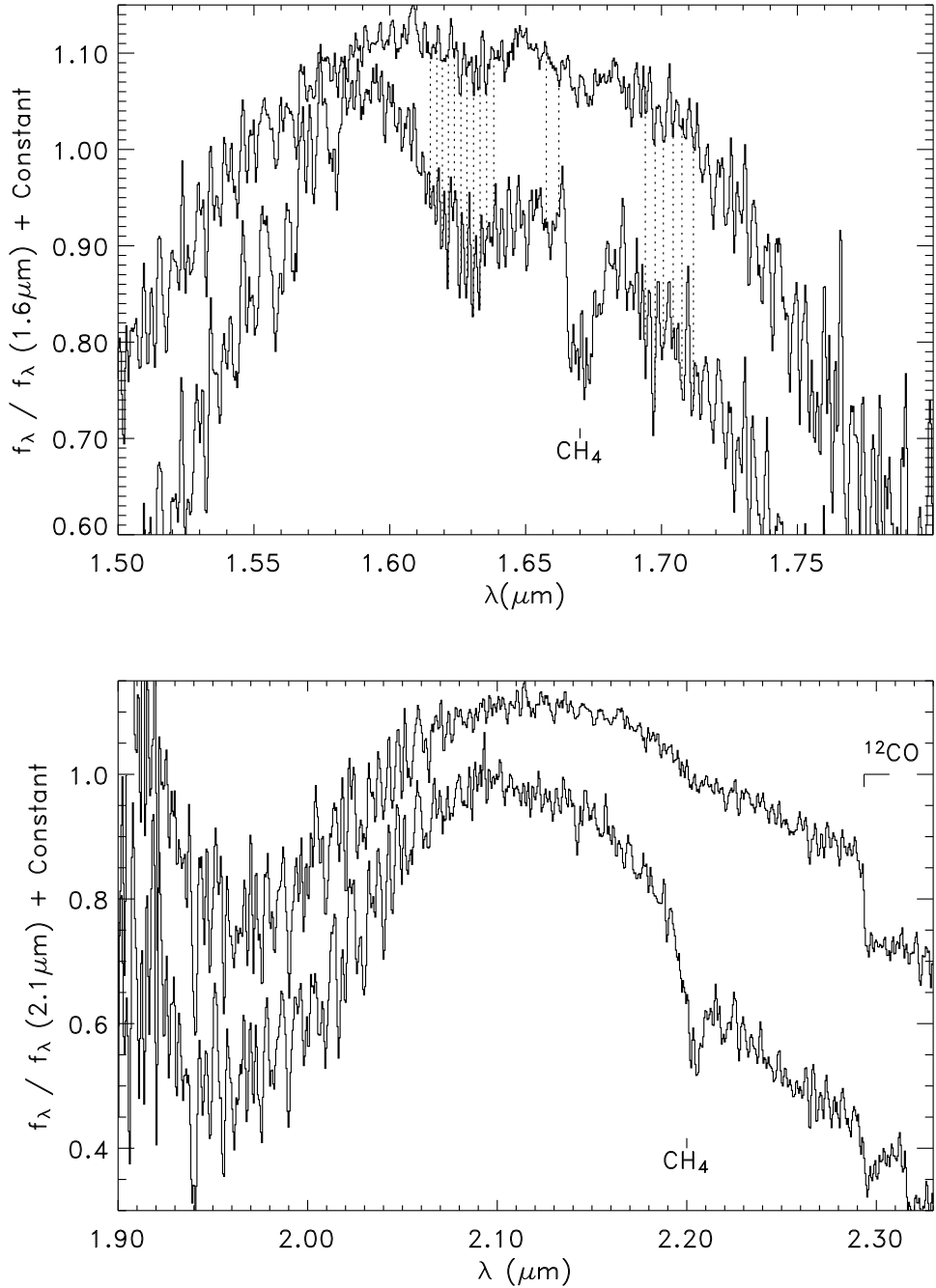}
\caption{\label{denis0255}The $H$- (upper panel) and $K$-band (lower panel) spectra of DENIS 0255$-$4700 (upper spectrum) and SDSS 1254$-$0122 (lower spectrum).  The CH$_4$ features identified in the spectrum of the 2MASS 0559$-$1404 are indicated with dotted lines.  Note not all of the CH$_4$ lines identified in the spectrum of 2MASS 0559$-$1404 are present in the spectra of DENIS 0255$-$4700 and SDSS 1254$-$0122.}

\end{figure}

DENIS 0255$-$4700 was classified as bdL6 by \citet{1999AJ....118.2466M} and reclassified as $\sim$L8 by \citet{2000AJ....120..447K}.  Figure \ref{denis0255} shows the $H$-band spectra of DENIS 0255$-$4700 and SDSS 1254$-$0122 (T2) in the upper panel and the $K$-band spectra of the same dwarfs in the lower panel.  In the $H$ band, the 2$\nu_3$ band of CH$_4$ at 1.67 $\mu$m is clearly present in the spectrum of DENIS 0255$-$4700.  In addition, the broad absorption trough of the 2$\nu_2 + \nu_3$ band centered at 1.63 $\mu$m, along with a number of individual CH$_4$ features (dotted lines), are also present in the spectrum of DENIS 0255$-$4700.  The $K$-band spectrum of DENIS 0255$-$4700 also shows evidence of CH$_4$ absorption.  There is an abrupt slope change in the $K$-band spectrum at $\sim$2.2 $\mu$m that is coincident with the absorption maximum of the $\nu_2 + \nu_3$ band.  This feature has been seen in other late-type L dwarfs but there has been some disagreement about its carrier.  For example, \citet{1999AJ....117.1010T} ascribe a similar feature in the spectrum of DENIS 0205$-$11AB to collision-induced H$_2$ absorption.  However, \citet{2001ApJ...561L.119N} find evidence of CH$_4$ absorption in both $H$- and $K$- band spectra of 2MASS 0902$+$3517 (L6.5).  Given the clear detection of CH$_4$ features in the $H$-band, we also ascribe this feature in the spectrum of DENIS 0255$-$4700 to CH$_4$.

Given the detection of CH$_4$ absorption in the $H$- and $K$-band spectra of DENIS 0255$-$4700, should it be reclassified as a T dwarf?  There are two spectral classification schemes currently in use for T dwarfs \citep{2002ApJ...564..421B,2002ApJ...564..466G}.  The defining characteristic of a T dwarf given by \citet{2002ApJ...564..466G} is ``... the appearance of methane absorption in the $H$ band, i.e., to its earliest appearance in \textit{both} the $H$ and $K$ bands.''  By this definition, DENIS 0255$-$4700 is a T dwarf.  However, we have computed the spectral indices for DENIS 0255$-$4700 defined by \citet{2002ApJ...564..466G} after smoothing the spectrum to $R$$\approx$500 (the resolving power of the spectral standards).  All of the \citeauthor{2002ApJ...564..466G} indices indicate DENIS 0255$-$4700 has a spectral type earlier than T0.  The discrepancy probably results from the difference in resolving power; the $H$-band methane features are simply too difficult to identify in $R\approx$500 spectrum.  Therefore, in the \citet{2002ApJ...564..466G} classification system, DENIS 0255$-$4700 should remain a late-type L dwarf.

Finally, \citet{2002ApJ...571L.151B} have suggested that late-type L and early-type T dwarfs may exhibit substantial photometric or spectroscopic variability due to the breakup of the cloud decks as they pass into the atmospheric convective zone.  As DENIS 0255$-$4700 is the brightest ($K=11.55$) late-type L dwarf, it would be an interesting target for both photometric and spectroscopic variability studies.

\section{Bolometric Fluxes and Luminosities} \label{Bolometric Fluxes and Luminosities}

The effective temperatures of ultracool dwarfs are typically determined by combining observationally measured bolometric luminosities with theoretical stellar radii \citep[e.g.,][]{1996ApJS..104..117L,2001ApJ...548..908L,2002ApJ...564..452L,2002AJ....124.1170D,2004AJ....127.3516G}.  The bolometric luminosities are measured using  absolutely flux-calibrated 1$-$2.5 $\mu$m spectra, $L'$-band photometry to account for the flux emitted between 2.5 and 3.6 $\mu$m, and a Rayleigh-Jeans tail at $\lambda >$ 3.6 $\mu$m.  Although atmospheric models indicate using $L'$-band photometry is a valid approximation, it has never been tested observationally.  Our spectra are ideal for testing this assumption.

In order to construct spectra suitable for integration over all wavelengths, we follow the standard practice of extending the spectra shortward of 0.6 $\mu$m using published optical photometry and longward of 4.1 $\mu$m using a Rayleigh-Jeans tail \citep{1987MNRAS.227..315B,1996ApJS..104..117L,2001ApJ...548..908L}.  The $U$, $B$, and $V$ magnitudes of the M dwarfs \citep{1992ApJS...82..351L} were converted to flux densities using the effective wavelength approach described by \citet{1996ApJS..104..117L}.  For each dwarf spectrum, we linearly interpolated from zero flux at zero wavelength, through the $U$-, $B$-, and $V$-band flux densities if available, to the flux density at the bluest wavelength of the spectrum.  The gaps in each spectrum at 1.85 and 2.6 $\mu$m were removed by linear interpolation between the flux densities at the gap edges.  Finally, we extended a Rayleigh-Jeans tail from the reddest wavelength of each spectrum to infinity.

Since there are limited observations of ultracool dwarfs at $\lambda > 4 \mu$m, we must resort to model atmospheres to determine the validity of assuming a Rayleigh-Jeans tail at these wavelengths.  We have compared the integrated flux of a Rayleigh-Jeans tail and the integrated flux of the NEXTGEN, DUSTY, and COND synthetic spectra \citep{2001ApJ...556..357A} at various effective temperatures at $\lambda > 4.1$ $\mu$m.  In gross terms, the NEXTGEN, DUSTY, and COND models are appropriate for M, L, and T dwarfs, respectively.  The integrated flux of a Rayleigh-Jeans tail at $\lambda > 4.1$ $\mu$m agrees with that of the models to $\sim$1\%, $\sim$5\%, and $\sim$20\% for M, L, and  T dwarfs, respectively.  The large discrepancy between the Rayleigh-Jeans tail and the models appropriate for the T dwarfs (COND) is a result of absorption bands of H$_2$O, CH$_4$, and NH$_3$ in the 6 to 12 $\mu$m region \citep{2004ApJS..154..418R}.

The integrated bolometric fluxes for the dwarfs with 0.6$-$4.1 $\mu$m spectra are listed in column 3 of Table \ref{FLbol}.  The errors in the bolometric fluxes include the error due to the photometry used to flux calibrate the spectra and an estimate of the systematic error introduced by assuming a Rayleigh-Jeans tail at $\lambda > 4.1$ $\mu$m.  \citet{1996ApJS..104..117L,2001ApJ...548..908L,2002ApJ...564..452L} have computed bolometric fluxes for eleven of the M and L dwarfs in our sample.  They compute these bolometric fluxes using near-infrared spectroscopy and $L'$-band photometry as described above.  Their bolometric flux measurements agree with our measurements within the errors for all eleven of the dwarfs.  This is not unexpected since the spectra are relatively featureless from 3.42 to 4.12 $\mu$m,  the wavelengths at which the MKO $L'$-band filter transmission is 50\% of the peak \citep{2002PASP..114..180T}.  The good agreement between the two methods means $L'$-band photometry can be used as a substitute for $L$-band spectroscopy greatly simplifying the calculation of the bolometric fluxes of M and L dwarfs.  The bolometric luminosities and absolute bolometric magnitudes of the dwarfs with measured parallaxes are also given in columns 5 and 6 of Table \ref{FLbol}.

In principle, we could now use the derived bolometric luminosities to determine the effective temperatures of the dwarfs in our sample.  However since our $L_{\mathrm{bol}}$ measurements agree with those of \citet{2001ApJ...548..908L}, our T$_{\mathrm{eff}}$ estimates will also agree and therefore we do not repeat this analysis.  We instead refer the reader to the recent work of \citet{2004AJ....127.3516G} that determined the effective temperatures of a large sample of ultracool dwarfs using the same technique as \citet{2001ApJ...548..908L}.

\section{Summary} \label{Summary}

We have presented a spectroscopic sequence of M, L, and T dwarfs, cover the wavelength range 0.6 to 4.1 $\mu$m, obtained with data from SpeX on the NASA IRTF, the IRCS on the Subaru Telescope, and previously published red-optical spectra.  Neutral lines of Al, Fe, Mg, Ca, and Ti are identified in the spectra of the M dwarfs but are all but absent in the spectra of the L dwarfs.  Nineteen new weak CH$_4$ absorption features are identified in the $H$-band spectrum of 2MASS 0559$-$1404.  We also confirm the presence of the 0$-$0 band of the $A$ $^4\Pi- X$ $^4\Sigma^-$ transition of VO at $\sim$1.06 $\mu$m in the spectra of late-type M dwarfs and detect it for the first time in the spectra of the early-type L dwarfs.  No absorption features of Cs, Rb, or CrH, hallmarks of L dwarf spectra in the red-optical, are found at $\lambda$ $>$ 1.0 $\mu$m.  The suite of spectroscopic indices derived by McLean et al. generally saturate for spectral types earlier than $\sim$M4 V thus limiting their usefulness to later spectral types.  The bolometric luminosities of the dwarfs in our sample agree with previously published results that make use of $L'$-band photometry rather than spectroscopy.  This results validates the use of $L'$-band photometry as a substitute for spectroscopy.  Finally, 2MASS 2224$-$0158 (L4.5) has a very red near-infrared spectrum and deep CO bands which may be indicative of low surface gravity and unusually thick condensate clouds.  The SpeX spectra can be obtained at the SpeX website \url{http://irtfweb.ifa.hawaii.edu/Facility/spex/} and the 0.6$-$4.1 $\mu$m spectra are available on request from the primary author.

\acknowledgments

The authors wish to thank Davy Kirkpatrick, Sandy Leggett, Adam Burgasser, Hugh Jones, Eduardo Mart{\'{\i}}n and Tod Henry and the RECONS team for providing the M, L, and T dwarf optical spectra, Sumner Davis and Ray Nassar for their high-resolution emission spectra of VO and CH$_4$, and Mark Marley and Richard Freedman for useful discussions.  We also thank the anonymous referee for his/her prompt review of the manuscript.  This publication makes use of data from the Two Micron All Sky Survey, which is a joint project of the University of Massachusetts and the Infrared Processing and Analysis Center, and funded by the National Aeronautics and Space Administration and the National Science Foundation.  This research has made use of the SIMBAD database, operated at CDS, Strasbourg, France.  M.C.C. acknowledges financial support from the NASA Infrared Telescope Facility and NASA through the Spitzer Space Telescope Fellowship Program.







\bibliographystyle{apj}
\bibliography{ref,tmp}

\clearpage




\clearpage

\clearpage

\clearpage

\clearpage

\clearpage

\clearpage

\clearpage

\clearpage

\clearpage

\clearpage

\clearpage

\clearpage

\clearpage

\clearpage

\begin{figure}
\includegraphics[width=5.5in]{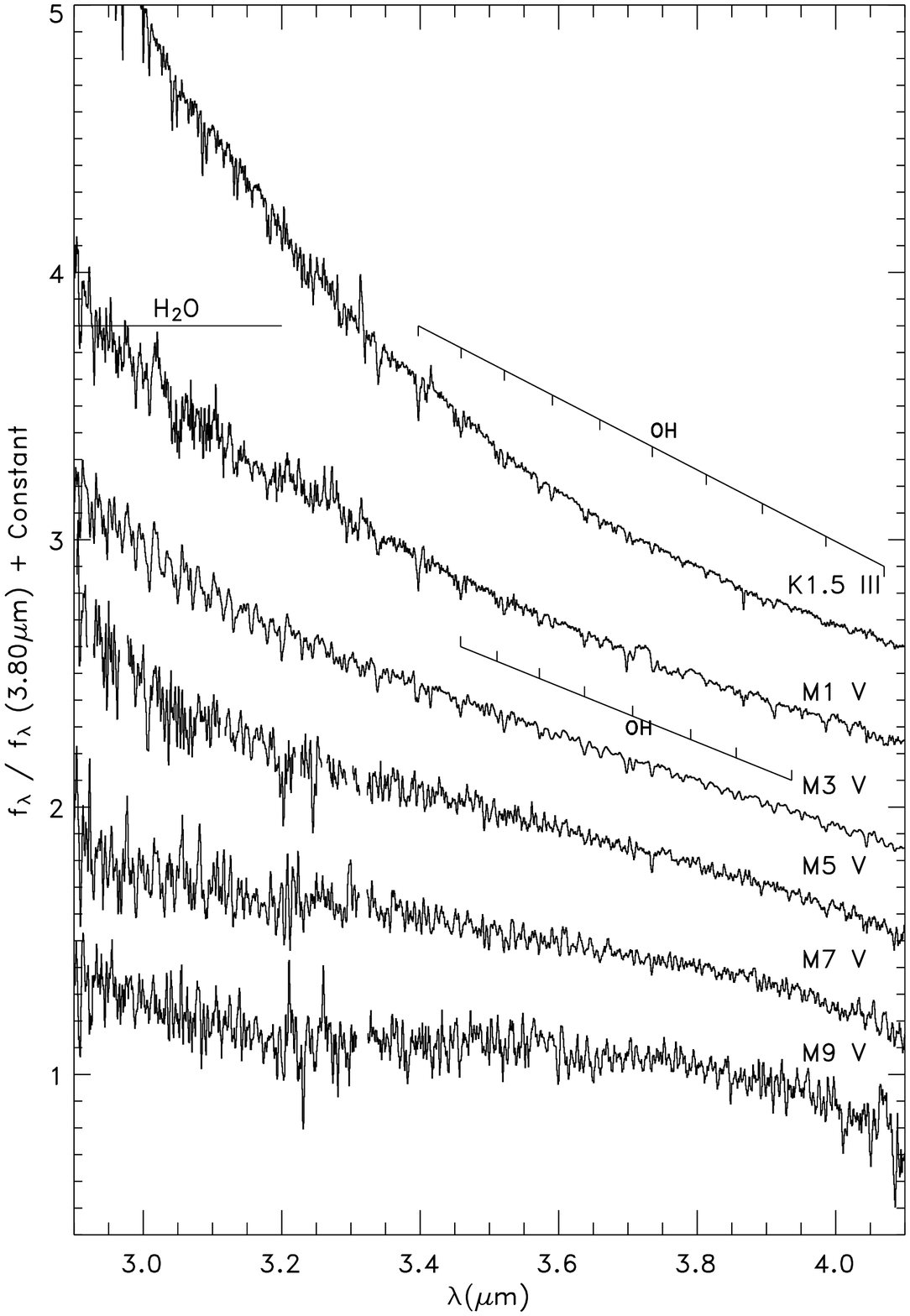}
\caption{\label{MLseq} $L$-band spectra of Arcturus (K1.5 III), Gl 229A (M1 V), Gl 388 (M3 V), Gl 51 (M5 V) vB 8 (M7 V) and LHS 2924 (M9 V).  The most prominent molecular and atomic features are indicated.} 
\end{figure}

\clearpage

\begin{figure}
\includegraphics[width=5.5in]{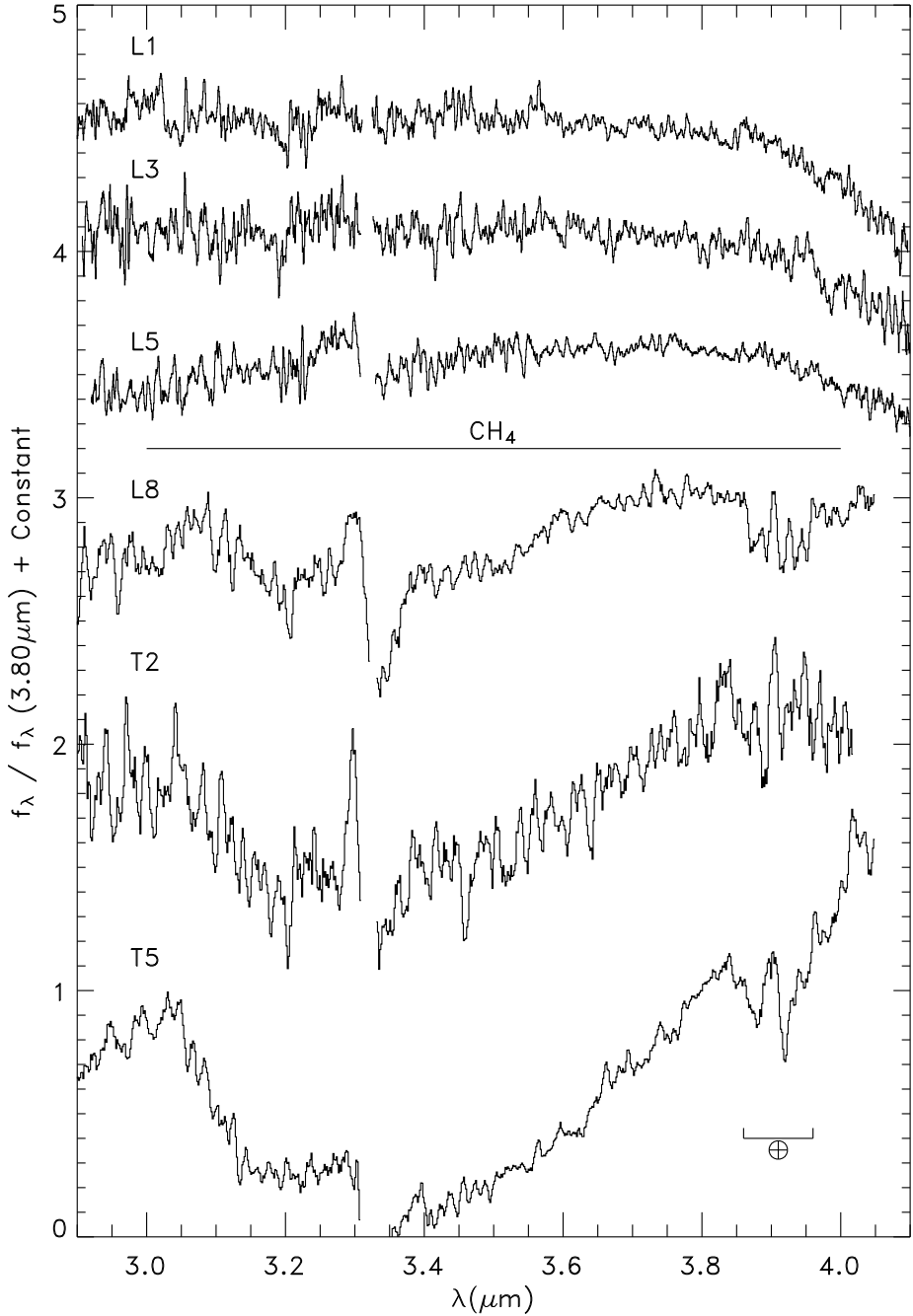}
\caption{\label{LTLseq} $L$-band spectra of 2MASS1439$+$1929 (L1), 2MASS 1506$+$1321 (L3), 2MASS 1507$-$1627 (L5), DENIS 0255$-$4700 (L8), SDSS 1254$-$0122 (T2), and 2MASS 0559$-$1404 (T5).  The most prominent molecular and atomic features are indicated.  The absorption features seen in the spectra of DENIS 0255$-$4700 and 2MASS 0559$-$1404 centered at $\sim$3.9 $\mu$m are due to incomplete removal of the N$_2$O telluric feature.}
\end{figure}

\clearpage

\clearpage

\clearpage

\clearpage

\clearpage

\clearpage


\input{tab1.tex}

\input{tab2.tex}

\clearpage

\input{tab3.tex}

\input{tab4.tex}

\input{tab5.tex}

\input{tab6.tex}

\input{tab7.tex}

\input{tab8.tex}

\input{tab9.tex}

\input{tab10.tex}

\end{document}

%% file: tab1.tex
\begin{deluxetable}{llr@{ $\pm$ }lr@{ $\pm$ }lr@{ $\pm$ }lr@{ $\pm$ }lr@{ $\pm$ }lc}
\tablecolumns{13}
\tabletypesize{\scriptsize} 
\tablewidth{0pc}
\tablecaption{\label{sample}The Sample}
\tablehead{
\colhead{Object} & 
\colhead{Spectral} & 
\multicolumn{2}{c}{{J}} &
\multicolumn{2}{c}{{H}} &
\multicolumn{2}{c}{{K}} &
\multicolumn{2}{c}{{L$'$}} &
\multicolumn{2}{c}{{$\pi$}} &
\colhead{$\pi$ Ref.} \\

\colhead{} & 
\colhead{Type} & 
\multicolumn{2}{c}{{(mag)}} &
\multicolumn{2}{c}{{(mag)}} &
\multicolumn{2}{c}{{(mag)}} &
\multicolumn{2}{c}{{(mag)}} &
\multicolumn{2}{c}{{(mas)}} &
\colhead{} \\

\colhead{(1)} & 
\colhead{(2)} & 
\multicolumn{2}{c}{{(3)}} &
\multicolumn{2}{c}{{(4)}} &
\multicolumn{2}{c}{{(5)}} &
\multicolumn{2}{c}{{(6)}} &
\multicolumn{2}{c}{{(7)}} &
\colhead{(8)}}

\startdata

Gl 229A                                      & M1 V     & 4.98    &  0.03   &  4.35     &  0.03   &  4.15    &  0.03   &  4.06   &  0.05                &  173.19  &  1.12                 & 1 \\
Gl 411                                       & M2 V     & 4.10    &  0.03   &  3.56     &  0.03   &  3.36    &  0.03   & \multicolumn{2}{c}{{$\cdots$}} &  392.52  &  0.91                 & 1 \\
Gl 388                                       & M3 V     & 5.449   &  0.027  &  4.843    &  0.020  &  4.593   &  0.017  & \multicolumn{2}{c}{{$\cdots$}} &  204.6   &  2.8                  & 2 \\
Gl 213                                       & M4 V     & 7.124   &  0.021  &  6.627    &  0.018  &  6.389   &  0.016  &  6.01   &  0.05                &  172.75  &  3.88                 & 1 \\
Gl 51                                        & M5 V     & 8.611   &  0.027  &  8.014    &  0.023  &  7.718   &  0.020  &  7.35   &  0.06                &  95.5    &  7.3                  & 2 \\
Gl 406                                       & M6 V     & 7.085   &  0.024  &  6.482    &  0.042  &  6.084   &  0.017  &  5.71   &  0.05                &  419.1   &  2.1                  & 2 \\
vB 8                                         & M7 V     & 9.776   &  0.029  &  9.201    &  0.024  &  8.816   &  0.023  & \multicolumn{2}{c}{{$\cdots$}} &  154.5   &  0.7                  & 3 \\
vB 10                                        & M8 V     & 9.90    &  0.03   &  9.24     &  0.03   &  8.80    &  0.03   &  8.18   &  0.05                &  170.25  &  1.37                 & 1 \\ 
LP 944$-$20                                  & M9 V     & 10.725  &  0.021  &  10.017   &  0.021  &  9.548   &  0.023  & \multicolumn{2}{c}{{$\cdots$}} &  201.4   &  4.2                  & 5 \\
LHS 2924                                     & M9 V     & 11.990  &  0.021  &  11.225   &  0.029  &  10.744  &  0.024  &  10.12  &  0.03                &  92.4    &  1.3                  & 2 \\
BRI 0021$-$0214                              & M9.5 V     & 11.992  &  0.035  &  11.084   &  0.022  &  10.539  &  0.023  &  9.78   &  0.13                &  86.6    &  4.0                  & 4 \\
2MASS J07464256$+$2000321AB                   & L0.5      & 11.759  &  0.020  &  11.007   &  0.022  &  10.468  &  0.022  &  9.67   &  0.03                &  81.9    &  0.3                  & 5 \\
2MASS J14392836$+$1929149                     & L1      & 12.759  &  0.019  &  12.041   &  0.019  &  11.546  &  0.022  &  10.80  &  0.05                &  69.6    &  0.5                  & 5 \\   
2MASS J02081833$+$2542533                     & L1      & 13.989  &  0.026  &  13.107   &  0.030  &  12.588  &  0.027  & \multicolumn{2}{c}{{$\cdots$}} &  \multicolumn{2}{c}{{$\cdots$}}  & $\cdots$ \\ 
Kelu-1                                       & L2      & 13.414  &  0.026  &  12.392   &  0.025  &  11.747  &  0.023  &  10.78  &  0.15                &  53.6    &  2.0                  & 5 \\     
2MASS J11463449$+$2230527AB                   & L3      & 14.165  &  0.028  &  13.182   &  0.026  &  12.590  &  0.026  & \multicolumn{2}{c}{{$\cdots$}} &  36.8    &  0.8                  & 5 \\   
2MASS J15065441$+$1321060                     & L3      & 13.365  &  0.023  &  12.380   &  0.021  &  11.741  &  0.019  & \multicolumn{2}{c}{{$\cdots$}} &  \multicolumn{2}{c}{{$\cdots$}}  & $\cdots$ \\
2MASS J00361617$+$1821104                     & L3.5      & 12.466  &  0.027  &  11.588   &  0.029  &  11.058  &  0.021  & 10.08   &  0.05                &  114.2   &  0.8                  & 5 \\
2MASS J22244381$-$0158521                     & L4.5      & 14.073  &  0.027  &  12.818   &  0.026  &  12.022  &  0.023  & 10.90   &  0.05                &  88.1    &  0.1                  & 5 \\   
2MASS J15074769$-$1627386                     & L5      & 12.830  &  0.027  &  11.895   &  0.024  &  11.312  &  0.026  & 9.98    &  0.03                &  136.4   &  0.6                  & 5 \\   
SDSS J053951.99$-$005902.0                    & L5      & 14.033  &  0.031  &  13.104   &  0.026  &  12.527  &  0.024  & 11.32   &  0.05                &  76.12   &  2.17                 & 6 \\
2MASS 15150083$+$4847416\tablenotemark{a}    & L6$^a$ & 14.111  &  0.029  &  13.099   &  0.031  &  12.500  &  0.024  & \multicolumn{2}{c}{{$\cdots$}} &  \multicolumn{2}{c}{{$\cdots$}}  & $\cdots$ \\ 
2MASS J08251968$+$2115521                     & L7.5      & 15.100  &  0.034  &  13.792   &  0.032  &  13.028  &  0.026  & 11.53   &  0.03                &  93.8    &  1.0                  & 5 \\   
DENIS-P 025503.3$-$470049.0                    & L8      & 13.246  &  0.027  &  12.204   &  0.024  &  11.558  &  0.024  & \multicolumn{2}{c}{{$\cdots$}} &  \multicolumn{2}{c}{{$\cdots$}}  & $\cdots$ \\
SDSS J125453.90$-$012247.4                    & T2      & 14.891  &  0.035  &  14.09    &  0.025  &  13.837  &  0.054  & 12.25   &  0.05                &  84.9    &  1.9                  & 5 \\   
2MASS J05591915$-$1404489                     & T5      & 13.802  &  0.024  &  13.679   &  0.044  &  13.577  &  0.052  & 12.14   &  0.05                &  97.7    &  1.3                  & 5 \\

\enddata

\tablecomments{$J$, $H$, and $K$-band photometry from the 2MASS Point Source Catalog except for Gl 229A, Gl 411, and vB 10 which are from \citet{1992ApJS...82..351L} and on the CIT system.  $L'$-band photometry from \citet{1998ApJ...509..836L}, \citet{2002ApJ...564..452L}, \citet{2004AJ....127.3516G}.}

\tablerefs{(1) \citet{1997A&A...323L..49P}; (2) \citet{1995gcts.book.....V}; (3) \citet{1992AJ....103..638M}; (4) \citet{1995AJ....110.3014T}; (5) \citet{2002AJ....124.1170D}; (6) \citet{2004AJ....127.2948V}}

\tablenotemark{a}{Infrared spectral type}

\end{deluxetable}

%% file: tab2.tex
\begin{deluxetable}{llclccl}
\tablecolumns{7}
\tabletypesize{\scriptsize}
\tablewidth{0pc}
\tablecaption{\label{spexlog}Log of SpeX Observations}
\tablehead{
\colhead{Object} & 
\colhead{Spectral} & 
\colhead{UT Date} & 
\colhead{Spectroscopy} & 
\colhead{$R$} & 
\colhead{Exp. Time} & 
\colhead {A0 V} \\

\colhead{} & 
\colhead{Type} & 
\colhead{} & 
\colhead{Mode} & 
\colhead{} & 
\colhead{(sec)} & 
\colhead{Standard} \\

\colhead{(1)} & 
\colhead{(2)} & 
\colhead{(3)} & 
\colhead{(4)} & 
\colhead{(5)} & 
\colhead{(6)} & 
\colhead {(7)}}

\startdata

Gl 229A             & M1 V   & 2002-01-10 & SXD      & 2000    & \phn200    & HD 42301 \\
                    &        & 2002-01-11 & LXD2.1   & 2500    & \phn250    & HD 42301 \\
Gl 411              & M2 V   & 2000-12-10 & SXD      & 2000    & \phn\phn60 & HD 88960 \\
                    &        & 2000-12-10 & LXD2.3   & 2500    & \phn500    & HD 88960 \\
Gl 388              & M3 V   & 2000-12-10 & SXD      & 2000    & \phn180    & HD 88960 \\
                    &        & 2000-12-10 & LXD2.3   & 1500    & \phn250    & HD 88960 \\
Gl 213              & M4 V   & 2000-12-09 & SXD      & 2000    & \phn720    & HD 34203 \\
                    &        & 2000-12-09 & LXD2.3   & 2500    & \phn500    & HD 34203 \\
Gl 51               & M5 V   & 2000-11-06 & SXD      & 2000    & \phn600    & HD 5071 \\
                    &        & 2000-11-06 & LXD2.3   & 1500    & \phn250    & HD 11946 \\
Gl 406              & M6 V   & 2001-01-25 & SXD      & 2000    & \phn360    & HD 97585 \\
                    &        & 2001-01-24 & LXD2.3   & 2500    & \phn500    & HD 97585 \\
vB 8                & M7 V   & 2001-07-12 & SXD      & 2000    & 1440       & HD 148968 \\
                    &        & 2001-07-12 & LXD1.9   & \phn938 & \phn660    & HD 148968 \\
vB 10               & M8 V   & 2001-07-13 & SXD      & 2000    & 1440       & HD 183324 \\
                    &        & 2001-07-13 & LXD1.9   & \phn938 & \phn600    & HD 183324 \\
LP 944$-$20         & M9 V   & 2001-01-24 & SXD      & 2000    & 1080       & HD 18735 \\ 
                    &        & 2002-01-11 & LXD1.9   & \phn938 & 2400       & HD 183324 \\
LHS 2924            & M9 V   & 2003-02-23 & SXD      & 2000    & 1200       & HD 127304 \\
                    &        & 2003-02-24 & LXD1.9   & \phn938 & \phn600    & HD 127304 \\
BRI 0021$-$0214     & M9.5 V & 2001-10-12 & SXD      & 2000    & 1440       & HD 9485   \\
                    &        & 2000-11-06 & LXD1.9   & \phn938 & \phn720    & HD 1663   \\
2MASS 0746$+$2000AB & L0.5   & 2001-01-25 & SXD      & 2000    & 1920       & HD 64648  \\
                    &        & 2002-02-28 & LXD1.9   & \phn938 & 1800       & HD 64648  \\
2MASS 1439$+$1929   & L1     & 2001-03-22 & SXD      & 2000    & 2160       & HD 131951 \\
                    &        & 2003-07-06 & LXD1.9   & \phn938 & 2280       & HD 131951 \\
2MASS 0208$+$2542   & L1     & 2002-11-11 & SXD      & 2000    & 2640       & HD 13869  \\
Kelu$-$1            & L2     & 2001-01-25 & SXD      & 2000    & 3120       & HD 119752 \\
                    &        & 2001-01-24 & LXD1.9   & \phn938 & 2460       & HD 119752 \\
                    &        & 2002-02-28 & LXD1.9   & \phn938 & 2280       & HD 119752 \\
2MASS 1146$+$2230AB & L3     & 2001-03-13 & SXD      & 1200    & 3360       & HD 105388 \\
                    &        & 2001-03-14 & SXD      & 1200    & 4800       & HD 105388 \\
2MASS 1506$+$1321   & L3     & 2001-03-13 & SXD      & 2000    & 2640       & HD 131951 \\                         
                    &        & 2002-02-28 & LXD1.9   & \phn938 & 1800       & HD 131951 \\
2MASS 0036$+$1821   & L3.5   & 2003-08-05 & SXD      & 2000    & 1200       & HD 6457   \\
                    &        & 2003-08-05 & LXD1.9   & \phn938 & 1800       & HD 6457   \\
2MASS 2224$-$0158   & L4.5   & 2001-08-07 & SXD      & 2000    & 4320       & HD 212404 \\
2MASS 1507$-$1627   & L5     & 2001-03-14 & SXD      & 2000    & 2160       & HD 133772 \\
                    &        & 2001-03-22 & SXD      & 2000    & 2400       & HD 133772 \\
                    &        & 2002-02-28 & LXD1.9   & \phn938 & 1200       & HD 124683 \\
SDSS 0539$-$0059    & L5     & 2001-01-24 & SXD      & 1200    & 2400       & HD 35656  \\
2MASS 1515$+$4847   & L6     & 2002-05-28 & SXD      & 2000    & 4800       & HD 116405 \\
2MASS 0825$+$2115   & L7.5   & 2001-03-12 & SXD      & 1200    & 6960       & HD 64648  \\
                    &        & 2001-03-13 & SXD      & 1200    & 2880       & HD 64648  \\
                    &        & 2001-03-14 & SXD      & 1200    & 3120       & HD 64648  \\
DENIS 0255$-$4700   & L8     & 2003-09-21 & SXD      & 2000    & 1500       & HD 21638  \\
SDSS 1254$-$0122    & T2     & 2001-01-25 & SXD      & 1200    & 1920       & HD 111744 \\
                    &        & 2001-03-13 & SXD      & 1200    & 4080       & HD 109309 \\
2MASS 0559$-$1404   & T5     & 2001-01-25 & SXD      & 1200    & 2880       & HD 41695  \\

\enddata
\end{deluxetable}

%% file: tab3.tex
\begin{deluxetable}{ccc}
\tablecolumns{3}
\tabletypesize{\scriptsize}
\tablewidth{0pc}
\tablecaption{\label{residuals}Dwarf Color Residuals}
\tablehead{
\colhead{$<$ $\delta_{J-H}$$>$} &
\colhead{$<$ $\delta_{H-K}$$>$} & 
\colhead{$<$ $\delta_{K-L'}$$>$}}

\startdata

$+$0.00 $\pm$ 0.04 & $+$0.02 $\pm$ 0.03 & $-$0.01 $\pm$ 0.07 \\

\enddata

\tablecomments{$\delta_{X-Y} = (X-Y)_{\mathrm{obs}} - (X-Y)_{\mathrm{synt}}$.  The errors are given by the RMS deviation.}

\end{deluxetable}

%% file: tab4.tex
\begin{deluxetable}{llcccl}
\tablecolumns{6}
\tabletypesize{\scriptsize}
\tablewidth{0pc}
\tablecaption{\label{ircslog}Log of IRCS Observations}
\tablehead{
\colhead{Object} & 
\colhead{Spectral} & 
\colhead{UT Date} & 
\colhead{$R$} & 
\colhead{Exp. Time} & 
\colhead {A0 V} \\

\colhead{} & 
\colhead{Type} & 
\colhead{} & 
\colhead{} & 
\colhead{(sec)} & 
\colhead{Standard} \\

\colhead{(1)} & 
\colhead{(2)} & 
\colhead{(3)} & 
\colhead{(4)} & 
\colhead{(5)} & 
\colhead {(6)}}

\startdata

2MASS 2224$-$0158 & L4.5  & 2002-12-12 & 425  & \phn560  & HD 215143 \\ 
SDSS 0539$-$0059  & L5  & 2002-12-09 & 425  & \phn600  & HD 37887 \\
2MASS 0825$+$2115 & L7.5  & 2001-05-06 & 212  & 1800  & HD 64648 \\
DENIS0255$-$4700  & L8  & 2002-12-09 & 212  & \phn252  & HD 28813 \\
SDSS 1254$-$0122  & T2  & 2001-05-06 & 212  & 1800  & HD 109309 \\
2MASS 0559$-$1404 & T5  & 2002-12-09 & 212  & 2590  & HD 47596 \\

\enddata
\end{deluxetable}

%% file: tab5.tex
\begin{deluxetable}{lccc}  
\tablecolumns{14}
\tabletypesize{\scriptsize} 
\tablewidth{0pc}
\tablecaption{\label{molfeat}Molecular Features In Dwarf Spectra}
\tablehead{
\colhead{Wavelength ($\mu$m)} &
\colhead{Species} &
\colhead{Transition} &
\colhead{Ref.} \\

\colhead{(1)} &
\colhead{(2)} &
\colhead{(3)} &
\colhead{(4)}} 

\startdata

0.89$-$0.99       & H$_2$O    & 3$\nu_3$,$\nu_1 +2\nu_3$, 2$\nu_1 + \nu_3$, 3$\nu_1$,$\nu_1 + 2\nu_2 +\nu_3$, 2$\nu_1 + 2\nu_2$ & 1 \\
0.9896 head       & FeH       & 0$-$0 band of $F$ $^4\Delta$ $-$ $X$ $^4\Delta$                                                 & 3 \\    
1.00606 Q-branch  & FeH       & 0$-$0 band of $F$ $^4\Delta$ $-$ $X$ $^4\Delta$                                     & 6 \\
1.05$-$1.08       & VO        & 0$-$0 band of $A$ $^4\Pi$ $-$ $X$ $^4\Sigma^-$                                                  & 5 \\   
1.09$-$1.20       & H$_2$O    & $\nu_2+2\nu_3$, $\nu_1 + \nu_2 + \nu_3$, 2$\nu_1 + \nu_2$, 3$\nu_2 +\nu_3$, $\nu_1 + 2\nu_2$    & 1 \\ 
1.1$-$1.24        & CH$_4$    & 3$\nu_3$                                                                                        & 9 \\
1.1939 head       & FeH       & 0$-$1 band of $F$ $^4\Delta$ $-$ $X$ $^4\Delta$                                                 & 3 \\       
1.22210 Q-branch  & FeH       & 0$-$0 band of $F$ $^4\Delta$ $-$ $X$ $^4\Delta$                                     & 6 \\
1.2389 head       & FeH       & 1$-$2 band of $F$ $^4\Delta$ $-$ $X$ $^4\Delta$                                                 & 3 \\       
1.3$-$1.51        & H$_2$O    & 2$\nu_3$, $\nu_1 + \nu_2$, 2$\nu_1$, 2$\nu_2 + \nu_3$, $\nu_1 + 2\nu_2$                         & 1 \\
1.58263 head      & FeH       & 0$-$0 band of $E$ $^4\Pi$ $-$ $A$ $^4\Pi$                                                       & 4 \\
1.59188 head      & FeH       & 0$-$0 band of $E$ $^4\Pi$ $-$ $A$ $^4\Pi$                                                       & 4 \\
1.62457 head      & FeH       & 0$-$0 band of $E$ $^4\Pi$ $-$ $A$ $^4\Pi$                                                       & 4 \\
1.6$-$1.8         & CH$_4$    & 2$\nu_3$, 2$\nu_2 + \nu_3$                                                                      & 8 \\
1.75$-$2.05       & H$_2$O    & $\nu_2 + \nu_3$,$\nu_1 + \nu_2$,3$\nu_2$                                                        & 1 \\
2.3$-$3.2         & H$_2$O    & $\nu_1$,$\nu_3$,2$\nu_2$                                                                        & 1 \\
2.15$-$2.5        & CH$_4$    & $\nu_3 + \nu_4$, $\nu_2 + \nu_3$                                                                & 8 \\
2.29352 head      & $^{12}$CO & 2$-$0 band of $X$ $^1\Sigma^+$ $-$ $X$ $^1\Sigma^+$                                             & 2 \\
2.32266 head      & $^{12}$CO & 3$-$1 band of $X$ $^1\Sigma^+$ $-$ $X$ $^1\Sigma^+$                                             & 2 \\
2.34483 head      & $^{13}$CO & 2$-$0 band of $X$ $^1\Sigma^+$ $-$ $X$ $^1\Sigma^+$                                             & 2 \\
2.35246 head      & $^{12}$CO & 4$-$2 band of $X$ $^1\Sigma^+$ $-$ $X$ $^1\Sigma^+$                                             & 2 \\
2.38295 head      & $^{12}$CO & 5$-$3 band of $X$ $^1\Sigma^+$ $-$ $X$ $^1\Sigma^+$                                             & 2 \\
2.41414 head      & $^{12}$CO & 6$-$4 band of $X$ $^1\Sigma^+$ $-$ $X$ $^1\Sigma^+$                                             & 2 \\
1.8$-$2.8           & H$_2$     & CIA                                                                                             & 11 \\
3.4$-$4.2 heads   & OH        & $\Delta \nu=+1,+2$ bandheads                                                                    & 10 \\
3.0$-$3.8         & CH$_4$    & $\nu_3$                                                                                         & 7 \\

\enddata        

\tablerefs{(1) \citet{1967ApJS...14..171A}; (2) \citet{1994ApJS...95..535G}; (3) \citet{1987ApJS...65..721P}; (4) \citet{2001ApJ...559..424W}; (5) \citet{1982JMS...92..391C}; (6) \citet{2003ApJ...582.1066C}; (7) \citet{2000ApJ...541L..75N}; (8) \citet{2000ApJ...536L..35L}; 
(9) \citet{1966ApJ...143..949D}; 10 \citet{2002AJ....124.3393W}; (11) \citet{1997A&A...324..185B}}

\end{deluxetable}

%% file: tab6.tex
\begin{deluxetable}{lllllllllll}
\tablecolumns{11}
\tabletypesize{\scriptsize}
\tablewidth{0pc}
\tablecaption{\label{atoms}Atomic Features in Dwarf Spectra}
\tablehead{
\colhead{Vacuum} & 
\colhead{Elements} & 
\colhead{} & 
\colhead{Vacuum} &
\colhead{Elements} & 
\colhead{} &
\colhead{Vacuum} &
\colhead{Elements} & 
\colhead{} &
\colhead{Vacuum} &
\colhead{Elements} \\

\colhead{$\lambda$ ($\mu$m)} & 
\colhead{} & 
\colhead{} & 
\colhead{$\lambda$ ($\mu$m)} &
\colhead{} & 
\colhead{} &
\colhead{$\lambda$ ($\mu$m)} &
\colhead{} & 
\colhead{} &
\colhead{$\lambda$ ($\mu$m)} &
\colhead{}}


\startdata

0.95500460  & Ti I        & &  1.0608472  &  Si I, Ti I  & &  1.2528860  & K I        & &  1.9781980     & Ca I        \\
0.96042912  & Ti I        & &  1.0664966  &  Ti I, Si I  & &  1.2682125  & Na I       & &  1.9817054     & Ca I        \\
0.96408301  & Ti I        & &  1.0697268  &  Si I        & &  1.2827300  & H I, Ti I  & &  1.9865941     & Ca I        \\
0.96516997  & Ti I        & &  1.1258730  &  Al I        & &  1.2851494  & Ti I       & &  1.9930196     & Ca I        \\
0.96774343  & Ti I        & &  1.1383850  &  Na I        & &  1.2883723  & Fe I       & &  1.9968176     & Ca I        \\
0.96923371  & Ti I        & &  1.1408517  &  Na I        & &  1.2904632  & Mn I       & &  2.1066365     & Mg I        \\
0.97085682  & Ti I        & &  1.1593413  &  Fe I        & &  1.2982085  & Mn I       & &  2.1098654     & Al I        \\
0.97315592  & Ti I        & &  1.1612996  &  Fe I        & &  1.3130514  & Al I       & &  2.1168456     & Al I, Si I  \\
0.97465017  & Ti I        & &  1.1640519  &  Fe I        & &  1.4878380  & Mg I       & &  2.1789074     & Si I, Ti I  \\
0.97708032  & Ti I        & &  1.1692427  &  K I, Fe I   & &  1.5028185  & Mg I       & &  2.2062996     & Na I        \\
0.97884568  & Ti I        & &  1.1778406  &  K I         & &  1.5046375  & Mg I       & &  2.2089866     & Na I        \\
0.98357543  & Ti I        & &  1.1801096  &  Fe I        & &  1.5167645  & K I        & &  2.2215927     & Ti I        \\
1.0219217   & Fe I        & &  1.1833522  &  Mg I        & &  1.5753070  & Mg I       & &  2.2237389     & Ti I        \\
1.0328393   & Sr II       & &  1.1891897  &  Fe I        & &  1.5771145  & Mg I       & &  2.2275054     & Fe I, Ti I  \\
1.0347329   & Fe I, Ca I  & &  1.1974243  &  Fe I, Ti I  & &  1.5894133  & Si I       & &  2.2317958     & Ti I        \\
1.0382309   & Si I        & &  1.1995492  &  Si I        & &  1.6723852  & Al I       & &  2.2390381     & Fe I        \\
1.0399835   & Fe I, Ti I  & &  1.2035899  &  Si I        & &  1.6755865  & Al I       & &  2.2626701     & Ca I        \\
1.0426753   & Fe I        & &  1.2086007  &  Mg I        & &  1.7110102  & Mg I       & &  2.2656055     & Ca I        \\
1.0471180   & Fe I        & &  1.2111957  &  Si I        & &  1.9314969  & Ca I       & &  2.2819887     & Mg I        \\
1.0500833   & Ti I        & &  1.2273618  &  Si I        & &  1.9458160  & Ca I       & &  2.3354050     & Na I        \\
1.0588292   & Ti I, Si I  & &  1.2436839  &  K I         & &  1.9509361  & Ca I       & &  2.3383542     & Na I        \\

\enddata
 
\end{deluxetable}

%% file: tab7.tex
\begin{deluxetable}{ccc}
\tablecolumns{2}
\tabletypesize{\scriptsize}
\tablewidth{0pc}
\tablecaption{\label{ch4features}CH$_4$ features in Dwarf Spectra}
\tablehead{
\colhead{Vacuum} & 
\colhead{Vacuum} \\

\colhead{Wavelength} & 
\colhead{Wavelength} \\

\colhead{($\mu$m)} & 
\colhead{($\mu$m)}}

\startdata

1.6149080 & 1.6382678  \\
1.6173155 & 1.6575994  \\
1.6193378 & 1.6622256  \\ 
1.6215579 & 1.6940340  \\
1.6237635 & 1.6976493  \\
1.6261759 & 1.7006882  \\
1.6283867 & 1.7042962  \\
1.6308079 & 1.7075190  \\
1.6332215 & 1.7117420  \\
1.6356431 & $\cdots$   \\

\enddata

\end{deluxetable}

%% file: tab8.tex
\begin{deluxetable}{llr@{ $\pm$ }lr@{ $\pm$ }lr@{ $\pm$ }lr@{ $\pm$ }l}
\tablecolumns{10}
\tabletypesize{\scriptsize}
\tablewidth{0pc}
\tablecaption{\label{KIew}K I Equivalent Widths} 
 
\tablehead{
\colhead{Object} &
\multicolumn{1}{c} {{Spectral}} &
\multicolumn{8}{c} {{EW (\AA)}} \\
\cline{3-4}
\cline{5-6}
\cline{7-8}
\cline{9-10}
 
\colhead{} &
\colhead{Type} &
\multicolumn{2}{c} {{1.169 $\mu$m}} &
\multicolumn{2}{c} {{1.177 $\mu$m}} &
\multicolumn{2}{c} {{1.244 $\mu$m}} &
\multicolumn{2}{c} {{1.253 $\mu$m}} \\
 
\colhead{(1)} &
\colhead{(2)} &
\multicolumn{2}{c} {{(3)}} &
\multicolumn{2}{c} {{(4)}} &
\multicolumn{2}{c} {{(5)}} &
\multicolumn{2}{c} {{(6)}}}
 
\startdata
 
Gl 229A & M1 V &  0.5 &  0.1 &  0.8 &  0.1 &  0.5 &  0.1 &  0.5 &  0.1 \\
Gl 411 & M2 V &  0.5 &  0.1 &  0.6 &  0.1 &  0.3 &  0.1 &  0.4 &  0.1 \\
Gl 388 & M3 V &  0.9 &  0.1 &  1.4 &  0.1 &  0.7 &  0.1 &  0.8 &  0.1 \\
Gl 213 & M4 V &  0.7 &  0.1 &  1.5 &  0.1 &  0.8 &  0.1 &  0.7 &  0.1 \\
Gl 51 & M5 V &  1.7 &  0.1 &  2.9 &  0.1 &  1.3 &  0.1 &  1.7 &  0.1 \\
Gl 406 & M6 V &  3.0 &  0.1 &  4.7 &  0.1 &  2.4 &  0.1 &  2.9 &  0.1 \\
vB 8 & M7 V &  4.2 &  0.1 &  6.4 &  0.2 &  3.2 &  0.2 &  4.3 &  0.1 \\
vB 10 & M8 V &  5.0 &  0.2 &  7.0 &  0.2 &  3.5 &  0.2 &  4.7 &  0.2 \\
LP 944$-$20 & M9 V &  5.5 &  0.3 &  7.9 &  0.2 &  3.8 &  0.3 &  5.2 &  0.3 \\
LHS 2924 & M9 V &  5.6 &  0.3 &  8.8 &  0.3 &  4.1 &  0.3 &  5.3 &  0.3 \\
BRI 0021$-$0214 & M9.5 V &  6.3 &  0.3 &  9.1 &  0.4 &  4.6 &  0.4 &  6.4 &  0.4 \\
2MASS 0746$+$2000AB & L0.5  &  6.7 &  0.3 & 10.2 &  0.4 &  4.8 &  0.4 &  6.6 &  0.4 \\
2MASS 1439$+$1929 & L1  &  6.9 &  0.4 & 10.5 &  0.4 &  4.9 &  0.4 &  7.3 &  0.5 \\
2MASS 0208$+$2542 & L1  &  6.4 &  0.5 & 10.4 &  0.7 &  3.8 &  0.6 &  6.3 &  0.5 \\
Kelu$-$1 & L2  &  6.3 &  0.4 &  9.1 &  0.5 &  3.8 &  0.5 &  6.4 &  0.5 \\
2MASS 1146$+$2230AB & L3  &  6.9 &  0.3 & 10.1 &  0.3 &  5.2 &  0.3 &  7.4 &  0.3 \\
2MASS 1506$+$1321 & L3  &  8.4 &  0.5 & 11.1 &  0.5 &  5.3 &  0.5 &  7.7 &  0.5 \\
2MASS 0036$+$1821 & L3.5  &  9.0 &  0.5 & 12.3 &  0.5 &  5.2 &  0.5 &  8.2 &  0.5 \\
2MASSS 2224$-$0158 & L4.5  &  6.7 &  0.8 & 12.0 &  0.6 &  4.7 &  0.4 &  6.5 &  0.5 \\
2MASS 1507$-$1627 & L5  &  8.6 &  0.5 & 11.8 &  0.4 &  4.1 &  0.4 &  7.8 &  0.5 \\
SDSS 0539$-$0059 & L5  &  8.8 &  0.4 & 12.1 &  0.4 &  4.0 &  0.4 &  8.3 &  0.5 \\
2MASS 1515$+$4847 & L6  &  6.9 &  0.7 & 10.5 &  0.5 &  3.4 &  0.4 &  6.2 &  0.4 \\
2MASS 0825$+$2115 & L7.5  &  4.4 &  0.3 &  5.1 &  0.2 &  2.8 &  0.1 &  3.9 &  0.2 \\
DENIS 0255$-$4700 & L8  &  5.8 &  0.6 &  7.0 &  0.3 &  2.1 &  0.2 &  3.9 &  0.2 \\
SDSS 1254$-$0122 & T2  &  5.6 &  0.7 &  9.3 &  0.4 &  3.9 &  0.3 &  6.8 &  0.4 \\
2MASS 0559$-$1404 & T5  &  8.7 &  1.0 & 13.2 &  0.9 &  3.7 &  0.3 &  8.4 &  0.5 \\
 
\enddata
\end{deluxetable}

%% file: tab9.tex
\begin{deluxetable}{llr@{ $\pm$ }lr@{ $\pm$ }lr@{ $\pm$ }lr@{ $\pm$ }lr@{ $\pm$ }lr@{ $\pm$ }lr@{ $\pm$ }l}
\tabletypesize{\scriptsize}
\tablecolumns{14}
\tablewidth{0pc}
\tablecaption{\label{metalsewt}Equivalent Widths of Other Lines}
\tablehead{
\colhead{Object} & 
\colhead{Spectral Type} & 
\multicolumn{12}{c}{{EW (\AA)}} \\

\cline{3-14}

\colhead{} & 
\colhead{} & 
\multicolumn{2}{c} {{Al I}} &
\multicolumn{2}{c} {{Mg I}} &
\multicolumn{2}{c} {{Fe I}} &
\multicolumn{2}{c} {{Al I}} &
\multicolumn{2}{c} {{Na I}} &
\multicolumn{2}{c} {{Ca I}} \\
 
\colhead{} & 
\colhead{} & 
\multicolumn{2}{c} {{1.125 $\mu$m}} &
\multicolumn{2}{c} {{1.183 $\mu$m}} &
\multicolumn{2}{c} {{1.189 $\mu$m}} &
\multicolumn{2}{c} {{1.132 $\mu$m}} &
\multicolumn{2}{c} {{2.207 $\mu$m}} &
\multicolumn{2}{c} {{2.264 $\mu$m}} \\

\colhead{(1)} &
\colhead{(2)} &
\multicolumn{2}{c} {{(3)}} &
\multicolumn{2}{c} {{(4)}} &
\multicolumn{2}{c} {{(5)}} &
\multicolumn{2}{c} {{(6)}} &
\multicolumn{2}{c} {{(7)}} &
\multicolumn{2}{c} {{(8)}}}

\startdata

Gl 229A             & M1 V &  0.7 &  0.1                    &  1.2 &  0.1                    &  1.0 &  0.1 &  2.9 &  0.1                    &  5.0 &  0.1                    &  4.7 &  0.1 \\ 
Gl 411              & M2 V &  0.6 &  0.1                    &  0.9 &  0.1                    &  0.6 &  0.1 &  2.4 &  0.1                    &  3.0 &  0.1                    &  2.7 &  0.1 \\ 
Gl 388              & M3 V &  0.4 &  0.1                    &  0.9 &  0.1                    &  1.2 &  0.1 &  2.6 &  0.1                    &  5.7 &  0.1                    &  5.0 &  0.1 \\ 
Gl 213              & M4 V &  0.5 &  0.1                    &  0.8 &  0.1                    &  0.9 &  0.1 &  2.5 &  0.1                    &  3.7 &  0.1                    &  2.5 &  0.1 \\ 
Gl 51               & M5 V &  0.1 &  0.1                    &  0.7 &  0.1                    &  1.3 &  0.1 &  2.6 &  0.1                    &  6.8 &  0.1                    &  3.5 &  0.2 \\ 
Gl 406              & M6 V & \multicolumn{2}{c}{{$\cdots$}} &  0.6 &  0.1                    &  1.1 &  0.1 &  2.3 &  0.2                    &  7.6 &  0.2                    &  2.0 &  0.2 \\ 
vB 8                & M7 V & \multicolumn{2}{c}{{$\cdots$}} &  0.6 &  0.1                    &  1.2 &  0.1 &  2.4 &  0.2                    &  4.9 &  0.3                    &  0.6 &  0.2 \\ 
vB 10               & M8 V & \multicolumn{2}{c}{{$\cdots$}} & \multicolumn{2}{c}{{$\cdots$}} &  0.8 &  0.2 &  2.0 &  0.3                    &  5.3 &  0.3                    & \multicolumn{2}{c}{{$\cdots$}} \\
LP 944$-$20         & M9 V & \multicolumn{2}{c}{{$\cdots$}} & \multicolumn{2}{c}{{$\cdots$}} &  0.9 &  0.2 &  1.0 &  0.4                    &  2.7 &  0.3                    & \multicolumn{2}{c}{{$\cdots$}} \\
LHS 2924            & M9 V & \multicolumn{2}{c}{{$\cdots$}} & \multicolumn{2}{c}{{$\cdots$}} &  0.9 &  0.2 &  1.4 &  0.4                    &  4.9 &  0.4                    & \multicolumn{2}{c}{{$\cdots$}} \\
BRI 0021$-$0214     & M9.5 V & \multicolumn{2}{c}{{$\cdots$}} & \multicolumn{2}{c}{{$\cdots$}} &  0.8 &  0.2 & \multicolumn{2}{c}{{$\cdots$}} &  2.8 &  0.4                    & \multicolumn{2}{c}{{$\cdots$}} \\
2MASS 0746$+$2000AB & L0.5  & \multicolumn{2}{c}{{$\cdots$}} & \multicolumn{2}{c}{{$\cdots$}} &  1.1 &  0.2 & \multicolumn{2}{c}{{$\cdots$}} &  2.0 &  0.3                    & \multicolumn{2}{c}{{$\cdots$}} \\
2MASS 1439$+$1929   & L1  & \multicolumn{2}{c}{{$\cdots$}} & \multicolumn{2}{c}{{$\cdots$}} &  1.0 &  0.2 & \multicolumn{2}{c}{{$\cdots$}} &  1.3 &  0.4                    & \multicolumn{2}{c}{{$\cdots$}} \\
2MASS 0208$+$2542   & L1  & \multicolumn{2}{c}{{$\cdots$}} & \multicolumn{2}{c}{{$\cdots$}} &  1.2 &  0.3 & \multicolumn{2}{c}{{$\cdots$}} & \multicolumn{2}{c}{{$\cdots$}} & \multicolumn{2}{c}{{$\cdots$}} \\
Kelu$-$1            & L2  & \multicolumn{2}{c}{{$\cdots$}} & \multicolumn{2}{c}{{$\cdots$}} &  0.8 &  0.2 & \multicolumn{2}{c}{{$\cdots$}} & \multicolumn{2}{c}{{$\cdots$}} & \multicolumn{2}{c}{{$\cdots$}} \\
2MASS 1146$+$2230AB & L3  & \multicolumn{2}{c}{{$\cdots$}} & \multicolumn{2}{c}{{$\cdots$}} &  0.6 &  0.2 & \multicolumn{2}{c}{{$\cdots$}} & \multicolumn{2}{c}{{$\cdots$}} & \multicolumn{2}{c}{{$\cdots$}} \\
2MASS 1506$+$1321   & L3  & \multicolumn{2}{c}{{$\cdots$}} & \multicolumn{2}{c}{{$\cdots$}} &  0.8 &  0.3 & \multicolumn{2}{c}{{$\cdots$}} & \multicolumn{2}{c}{{$\cdots$}} & \multicolumn{2}{c}{{$\cdots$}} \\
2MASS 0036$+$1821   & L3.5  & \multicolumn{2}{c}{{$\cdots$}} & \multicolumn{2}{c}{{$\cdots$}} &  1.0 &  0.3 & \multicolumn{2}{c}{{$\cdots$}} & \multicolumn{2}{c}{{$\cdots$}} & \multicolumn{2}{c}{{$\cdots$}} \\
2MASS 2224$-$0158   & L4.5  & \multicolumn{2}{c}{{$\cdots$}} & \multicolumn{2}{c}{{$\cdots$}} &  0.8 &  0.3 & \multicolumn{2}{c}{{$\cdots$}} & \multicolumn{2}{c}{{$\cdots$}} & \multicolumn{2}{c}{{$\cdots$}} \\

\enddata

\end{deluxetable}

%% file: tab10.tex
\begin{deluxetable}{llr@{ $\pm$ }lr@{ $\pm$ }lr@{ $\pm$ }lr@{ $\pm$ }l}
\tablecolumns{14}
\tabletypesize{\scriptsize}
\tablewidth{0pc}
\tablecaption{\label{FLbol}$f_{\mathrm{bol}}$ and $L_{\mathrm{bol}}$ of the Dwarfs}
\tablehead{
\colhead{Object} &
\colhead{Spectral} &
\multicolumn{2}{c} {{$f_{\mathrm{bol}}$}} &
\multicolumn{2}{c} {{$m_{\mathrm{bol}}$}\tablenotemark{a}} &
\multicolumn{2}{c} {{$\log_{10} (L_{\mathrm{bol}}/L_{\odot}$)\tablenotemark{b}}} &
\multicolumn{2}{c} {{$M_{\mathrm{bol}}$}} \\
 
\colhead{} &
\colhead{Type} &
\multicolumn{2}{c} {{(W m$^{-2}$)}} &
\multicolumn{2}{c} {{}} &
\multicolumn{2}{c} {{}} &
\multicolumn{2}{c} {{}} \\
 
\colhead{(1)} &
\colhead{(2)} &
\multicolumn{2}{c} {{(3)}} &
\multicolumn{2}{c} {{(4)}} &
\multicolumn{2}{c} {{(5)}} &
\multicolumn{2}{c} {{(6)}}}
 
\startdata
Gl 229A & M1 V & 5.19E$-$11 & 1.43E$-$12 &  6.72 &  0.03 & $-$1.27 &  0.01 &  7.92 &  0.03 \\
Gl 411  & M2 V & 1.10E$-$10 & 3.03E$-$12 &  5.91 &  0.03 & $-$1.66 &  0.01 &  8.88 &  0.03 \\
Gl 388  & M3 V & 3.24E$-$11 & 7.11E$-$13 &  7.24 &  0.03 & $-$1.62 &  0.02 &  8.79 & 0.04  \\      
Gl 213 & M4 V & 5.97E$-$12 & 1.24E$-$13 &  9.07 &  0.02 & $-$2.21 &  0.02 & 10.26 &  0.05 \\ 
Gl 51 & M5 V & 1.49E$-$12 & 3.49E$-$14 & 10.58 &  0.03 & $-$2.30 &  0.07 & 10.48 &  0.17 \\ 
Gl 406 & M6 V & 5.90E$-$12 & 1.40E$-$13 &  9.08 &  0.03 & $-$2.98 &  0.01 & 12.20 &  0.03 \\ 
vB 8 & M7 V & 4.73E$-$13 & 1.18E$-$14 & 11.83 &  0.03 & $-$3.21 &  0.01 & 12.77 &  0.03 \\ 
vB 10 & M8 V & 4.28E$-$13 & 1.19E$-$14 & 11.93 &  0.03 & $-$3.34 &  0.01 & 13.09 &  0.03 \\ 
LP 944$-$20 & M9 V & 2.02E$-$13 & 4.67E$-$15 & 12.75 &  0.03 & $-$3.81 &  0.02 & 14.27 &  0.05 \\ 
LHS 2924 & M9 V & 6.57E$-$14 & 1.62E$-$15 & 13.97 &  0.03 & $-$3.62 &  0.02 & 13.79 &  0.04 \\ 
BRI 0021$-$0214 & M9.5 V & 7.43E$-$14 & 1.88E$-$15 & 13.84 &  0.03 & $-$3.51 &  0.04 & 13.52 &  0.10 \\ 
2MASS 0746$+$2000AB & L0.5  & 8.07E$-$14 & 2.09E$-$15 & 13.75 &  0.03 & $-$3.43 &  0.01 & 13.31 &  0.03 \\ 
2MASS 1439$+$1929 & L1  & 3.00E$-$14 & 7.31E$-$16 & 14.82 &  0.03 & $-$3.72 &  0.01 & 14.03 &  0.03 \\ 
Kelu$-$1 & L2  & 2.02E$-$14 & 5.09E$-$16 & 15.25 &  0.03 & $-$3.66 &  0.03 & 13.89 &  0.09 \\ 
2MASS 1506$+$1321 & L3  & 2.21E$-$14 & 5.74E$-$16 & 15.15 &  0.03 & \multicolumn{2}{c}{{$\cdots$}} & \multicolumn{2}{c}{{$\cdots$}} \\
2MASS 0036$+$1821 & L3.5  & 4.55E$-$14 & 1.35E$-$15 & 14.37 &  0.03 & $-$3.97 &  0.01 & 14.65 &  0.04 \\ 
2MASSS 2224$-$0158 & L4.5  & 1.58E$-$14 & 4.84E$-$16 & 15.51 &  0.03 & $-$4.20 &  0.01 & 15.24 &  0.03 \\ 
2MASS 1507$-$1627 & L5  & 3.54E$-$14 & 1.08E$-$15 & 14.64 &  0.03 & $-$4.16 &  0.01 & 15.15 &  0.03 \\ 
SDSS 0539$-$0559 & L5  & 1.20E$-$14 & 3.79E$-$16 & 15.81 &  0.03 & $-$4.19 &  0.03 & 15.22 &  0.07 \\ 
2MASS 0825$+$2115 & L7  & 7.51E$-$15 & 3.51E$-$16 & 16.32 &  0.05 & $-$4.58 &  0.02 & 16.18 &  0.05 \\ 
DENIS 0255$-$4700 & L8  & 3.15E$-$14 & 1.46E$-$15 & 14.76 &  0.05 & \multicolumn{2}{c}{{$\cdots$}} & \multicolumn{2}{c}{{$\cdots$}} \\
SDSS 1254$-$0122 & T2  & 4.82E$-$15 & 3.30E$-$16 & 16.81 &  0.07 & $-$4.68 &  0.04 & 16.45 &  0.09 \\ 
2MASS 0559$-$1404 & T5  & 8.55E$-$15 & 6.22E$-$16 & 16.18 &  0.08 & $-$4.56 &  0.03 & 16.13 &  0.08 \\ 
 
\enddata
 
\tablenotemark{a}{$m_{\mathrm{bol}}  = -2.5 \times \log(f_{\mathrm{bol}}) - 18.988 \label{mbol}$ assuming $L_{\odot} = 3.86\times10^{26}$ W and $M_{\mathrm{bol}\odot} = +4.74$.}
 
\tablenotetext{b}{$\log(L_{\mathrm{bol}}/L_{\odot}) = \log(f_{\mathrm{bol}}) - 2\times \log(\pi) + 7.4913$ assuming $L_{\odot} = 3.86\times10^{26}$ W.}
 
\end{deluxetable}